\documentclass[superscriptaddress]{revtex4-2}

\usepackage{fouriernc}
\usepackage{graphicx}
\graphicspath{{figures}}
\usepackage[utf8]{inputenc}
\usepackage{multirow}
\usepackage[dvipsnames]{xcolor}
\usepackage{xspace}
\usepackage{amsmath}
\usepackage{amssymb}
\usepackage{physics}
\usepackage[dvipsnames]{xcolor}
\usepackage{hyperref}  
\usepackage{caption}
\usepackage{comment}
\usepackage{setspace}
\usepackage{bm}
\usepackage{float}
\usepackage[export]{adjustbox}
\usepackage{subcaption}
\usepackage{microtype}
\usepackage{orcidlink}

\definecolor{darkperiwinkle}{RGB}{102, 102, 128}

\definecolor{darkperiwinkle}{RGB}{102, 102, 128}

\newcommand{\newacronym}[3]{%
  \newcommand{#1}{#2 (#3)\xspace%
    \renewcommand{#1}{#3\xspace}%
  }%
}

\newacronym{\GRMHD}{General Relativistic Resistive Magnetohydrodynamics}{GR-RMHD}
\newacronym{\MHD}{Magnetohydrodynamics}{MHD}
\newacronym{\RHS}{right hand side}{RHS}
\newacronym{\ECP}{Exascale Computing Project}{ECP}

\begin{document}

\title{A resistive MHD module in the GPU-accelerated GRMHD code \texttt{GRaM-X}}

\author{Sara Azizi \orcidlink{0009-0004-6525-8712}}
\email{s.azizi@uva.nl}
\affiliation{Anton Pannekoek Institute for Astronomy and GRAPPA, 
University of Amsterdam, 
Science Park 904, 1098 XH Amsterdam, The Netherlands}

\author{Swapnil Shankar \orcidlink{0000-0002-5109-0929}}
\email{s.shankar@uva.nl}
\affiliation{Faculty of Mathematics, Informatics and Natural Sciences, University of Hamburg, Gojenbergsweg 112, 21029 Hamburg, Germany}
\affiliation{Anton Pannekoek Institute for Astronomy and GRAPPA, 
University of Amsterdam, 
Science Park 904, 1098 XH Amsterdam, The Netherlands}

\author{Philipp M\"osta \orcidlink{0000-0002-9371-1447}}
\email{p.moesta@uva.nl}
\affiliation{GRAPPA, 
Anton Pannekoek Institute for Astronomy and Institute of
High-Energy Physics, University of Amsterdam,
Science Park 904, 1098 XH Amsterdam, The Netherlands}

\author{Roland Haas \orcidlink{0000-0003-1424-6178}}
\affiliation{Department of Physics and Astronomy, University of British Columbia, Vancouver, British Columbia, Canada }
\affiliation{National Center for Supercomputing applications, University of Illinois, Urbana, Illinois, USA}
\affiliation{Department of Physics, University of Illinois, Urbana, Illinois, USA}

\author{Erik Schnetter \orcidlink{0000-0002-4518-9017}}
\affiliation{Perimeter Institute for Theoretical Physics, Waterloo, Ontario, Canada}
\affiliation{Department of Physics and Astronomy, University of Waterloo, Waterloo, Ontario, Canada}
\affiliation{Center for Computation \& Technology, Louisiana State University, Baton Rouge, Louisiana, USA}


\begin{abstract}
Relativistic macroscopic plasma dynamics can be described by general-relativistic magnetohydrodynamics. In many high-energy astrophysical settings, such as the interior dynamics of magnetized stars, the ideal GRMHD approximation, in which we assume infinite conductivity, provides an excellent description. However, ideal GRMHD neglects resistive effects that are essential for processes such as magnetic reconnection, dissipation, and magnetospheric dynamics. Incorporating resistivity into astrophysical plasma models accounts for the fact that plasmas in such environments are not perfect conductors. We present a resistive version of the GPU-accelerated GRMHD code \texttt{GRaM-X}, which evolves the full resistive GRMHD equations using the Z4c formalism for Einstein's equations. We implement a second-order implicit–explicit Runge–Kutta scheme to handle stiff source terms, obtain the primitive quantities from the conserved quantities using a one-dimensional recovery method,  and employ the HLLE Riemann solver in combination with TVD and WENO reconstruction schemes. We validate the module using a range of standard tests, including 1D shocktubes, current sheets, Alfv\'{e}n waves, 2D cylindrical explosions, and 3D TOV stars. The results of these tests demonstrate accurate recovery of the ideal MHD limit, correct resistive behavior, and stable evolution in dynamical spacetimes. Leveraging the GPU-accelerated resistive version of \texttt{GRaM-X} enables efficient large-scale simulations, paving the way for realistic studies of binary mergers, accretion flows, and relativistic jets within the framework of multi-messenger astrophysics.
\end{abstract}


\maketitle

\setlength{\parskip}{2ex}

\section{Introduction}
\label{sec:Introduction}

Magnetic fields are fundamental components of numerous astrophysical processes. These astrophysical phenomena frequently occur in high-energy environments, where the dynamics are simultaneously happening at both micro and macro scales. Magnetic fields play a crucial role in pulsars, magnetars, compact X-ray systems, supernovae, and gamma-ray bursts, as well as larger structures such as radio galaxies, quasars, and active galactic nuclei (AGN).
Precise modeling of the behavior of such systems requires solving the equations of magnetohydrodynamics (MHD) within the framework of general relativity (GR).
These equations describe a plasma that interacts with electromagnetic fields in a curved spacetime, that may be static or dynamic. 
GRMHD models can describe the behavior of astrophysical plasmas well, for instance, accretion onto compact objects as well as the dynamics of outflows from them on a global scale \cite{McKinney2006, Font2008, Mizuno2022}.

In most cases, the electrical conductivity of the plasma is sufficiently high that the ideal MHD approximation can be used. In this approximation, the electrical resistivity is considered negligible, resulting in the magnetic field being solely advected by the fluid motion. This assumption significantly simplifies the equations and modeling and has been the basis for the development of many existing numerical codes for astrophysical simulations~\cite{komissarov1999, DelZanna2003, DelZanna2007, Giacomazzo2007, Farris2008, farris2011binary, Bucciantini2011, zink2011horizon, Msta2013, Palenzuela2013, Etienne2015, Ciolfi2019, Mewes2020, Mizuno2022, Kiuchi2022, Shankar2023}.
For example, in the dense interior of neutron stars, the properties of electrical and thermal transport are mainly determined by electrons, because they are primarily responsible for carrying electrical charge and thermal energy at high densities. At temperatures exceeding the crystallization temperature of the ions, the electrical and thermal conductivities are determined by electrons scattering off the ions. 
Even if the length scale of variation of the magnetic field is small relative to the star (e.g., on a length scale about one-tenth the radius of the star), the timescale for this magnetic field to change as a result of Ohmic diffusion is estimated to be several million years~\cite{Palenzuela2009}. 
As a result, under these temperatures and densities, processes occurring within the neutron star on significantly shorter timescales ($\lesssim$ a few seconds) remain largely unaffected by this form of diffusion. It is therefore reasonable to assume that the electrical conductivity in this medium is infinite, allowing for the application of the ideal MHD approximation to investigate these phenomena.

However, the ideal MHD approximation fails in extremely high-energy environments where a finite electrical resistivity is crucial. Scenarios such as compact-object mergers (neutron star-neutron star or neutron star-black hole) or regions near accretion disks surrounding AGNs produce plasmas with relatively low densities ($\rho \sim 10^{8{-}10}\,\mathrm{g\,cm^{-3}}$) compared to neutron star interiors, and significantly higher temperatures ($T \sim 10^{11{-}13}\,\mathrm{K}$)~\cite{Palenzuela2009}. These conditions influence the transport properties of matter, and resistive phenomena are expected to play a vital role in plasma dynamics and energy dissipation.
Moreover, plasma instabilities and magnetic reconnection can trigger particle acceleration and energy dissipation, giving rise to energetic electromagnetic flares~\cite{Ripperda2019}.
Magnetospheres around compact objects are highly magnetized and have low densities. In these regions, magnetic pressure dominates over gas pressure and leads to a force-free configuration where the fields adjust to cancel out any magnetic stresses. This regime is fundamentally different from ideal MHD. However, neither the ideal MHD nor the force-free equations can simultaneously describe both the dense interior and the magnetically dominated exterior of a compact object~\cite{Komissarov2002, McKinney2006, moesta2012detectability, Alic2012}.

The framework of general-relativistic resistive magnetohydrodynamics (GR-RMHD) allows a precise and systematic study of non-ideal effects in these systems~\cite{lichnerowicz1967, carter1991,Palenzuela2009, Dionysopoulou2013, Ripperda2019}. In the GRMHD framework, the general-relativistic Maxwell's equations are solved alongside the general-relativistic fluid equations. The key difference between GR-RMHD and ideal GRMHD is that, in the ideal case, the electric field is determined algebraically from the magnetic field and fluid velocity; however, in GR-RMHD, the electric and magnetic fields evolve independently.  
The GR-RMHD framework offers several key advantages. Most importantly, it provides a unified framework capable of simultaneously describing regions of 1) very high conductivity (like the interiors of compact objects, reducing to ideal MHD) and 2) low conductivity or vacuum regions (like the highly magnetized magnetospheres surrounding a compact object, which approach the Force-Free Electrodynamics regime). This inherent capability allows for the systematic study and distinction between physical resistivity and numerical resistivity, enabling more precise modeling of plasma properties~\cite{Palenzuela2009, Dionysopoulou2013}.

The introduction of resistivity (finite conductivity) in the GR-RMHD equations leads to significant numerical challenges. The electric field evolution in cases where the electrical resistivity is small but nonzero takes place on a substantially shorter timescale compared to the GRMHD evolution. This difference leads to the appearance of a stiff source term in Ampere's law, making the use of explicit numerical methods problematic in terms of stability and efficiency.
One of the key aspects of simulations within the GRMHD framework is the recovery of primitive variables from conservative variables. In the case of GR-RMHD, this transformation is more complicated due to the presence of a stiff electric field. A range of one-, two-, three-, and four-dimensional transformation methods have been proposed in~\cite{Palenzuela2009, Dionysopoulou2013,Palenzuela2013, Qian2016, Ripperda2019}, along with more recent advances in numerical methods aimed at overcoming the stiffness problem in the RMHD equations. One of the earliest efforts was made by~\cite{Komissarov2007}, in which a formulation of special-relativistic resistive magnetohydrodynamics (SR-RMHD) was developed using Strang time-splitting. 
Later, \cite{Takamoto2011} employed an implicit inversion method to handle the stiff source term, thereby relaxing the restrictive timestep constraint in SR-RMHD.
The first application of the implicit-explicit Runge–Kutta (IMEX) scheme to SR-RMHD was presented by \cite{Palenzuela2009}, where the stiff source terms are treated implicitly, while the non-stiff components such as those arising in ideal relativistic MHD are handled explicitly.
Building on this approach, \cite{Bucciantini2012} and \cite{Dionysopoulou2013} extended the IMEX method to GR-RMHD by incorporating the full resistive form of Ohm’s law.
IMEX methods offer a significant advantage in both SR-RMHD and GR-RMHD simulations: they alleviate the restrictive timestep constraints imposed by stiffness, without incurring the severe slowdowns typically associated with fully implicit schemes.

In this work, we present a new GPU-accelerated dynamical-spacetime GR-RMHD module in the code \texttt{GRaM-X}. We have incorporated several IMEX-RK Strong Stability Preserving (SSP) schemes, including IMEX–SSP2 $(2, 2, 2)$, IMEX–SSP2 $(3, 3, 2)$, IMEX–SSP3 $(3, 3, 2)$, and IMEX-SSP3 $(4, 3, 3)$~\cite{Palenzuela2009, conde2017implicit}. For simplicity, we adopt the IMEX-RK SSP2 $(2, 2, 2)$ scheme and use a one-dimensional primitive variables recovery technique to solve the GR-RMHD equations and recover primitive variables efficiently and stably, following the approach of~\cite{Palenzuela2009}.
This work represents our initial approach, and in the future, we plan to extend it to improve both performance and generality. The outline of the paper is as follows. In section~\ref{sec:Evolution_equations}, we introduce the full set of GR-RMHD equations, which includes a brief introduction to the 3+1 decomposition of general relativity, as well as Maxwell's equations and the MHD equations. We describe the numerical methods used to solve the GR-RMHD system in section~\ref{sec:Numerical_methods}. This includes an overview of the IMEX-RK schemes, their application to the resistive MHD equations, and the procedure for recovering primitive variables from conserved variables. In section~\ref{sec:Numerical_Tests}, we present a series of numerical tests in one, two, and three dimensions, covering a range of conductivity models. Finally, we summarize our conclusions and outline possible directions for future development in section~\ref{sec:Conclusion_sec}.

\section{Evolution equations}
\label{sec:Evolution_equations}

To comprehend how a fluid behaves in the presence of electromagnetic fields, we must examine three important sets of equations. The electromagnetic fields themselves are described in one set, the fluid's dynamics are the subject of another, and the third connects the two through their interactions. Maxwell’s equations govern the electromagnetic side, and energy and momentum conservation laws govern the fluid’s evolution. Ohm's law, which establishes the strength of the mutual influence between the fields and the fluid and depends on the fluid's microscopic properties, couples the two.

In the next sections, we will go through each of these equation sets one by one. Then we will show how combining them gives us the full set of GR-RMHD equations. Under certain conditions, this full system simplifies to more familiar cases like ideal MHD or the vacuum version of Maxwell’s equations.
In this work, we will use a spacetime metric with the signature \((- + + +)\) and adopt natural units so that the speed of light \( c = 1 \), and the gravitational constant \( G = 1 \). Greek indices will refer to spacetime coordinates (0 to 3), while Latin indices will be used just for spatial ones (1 to 3).

\subsection{3+1 Decomposition of General Relativity}

The 3+1 formalism is a useful strategy for rewriting the equations in an appropriate format when numerically solving GR-RMHD equations~\cite{rezzolla2013relativistic}. The $3 + 1$ formalism's line element, which depicts the motion of coordinate lines as seen by an Eulerian observer traveling with four-velocities, is written as follows:

\begin{equation}
ds^2 = -\alpha^2 dt^2 + \gamma_{ij} \left( dx^i + \beta^i dt \right) \left( dx^j + \beta^j dt \right),
\label{eq:3plus1_metric}
\end{equation}

\noindent and,

\begin{equation}
n_\mu = (-\alpha, 0, 0, 0), \qquad
n^\mu = \left( \frac{1}{\alpha}, -\frac{\beta^i}{\alpha} \right),
\label{eq:normal_vector}
\end{equation}

\noindent where the lapse function is represented by $\alpha$, and the shift three-vector by $\beta^i$. The metric tensor and its inverse are expressed as follows:

\begin{equation}
g_{\mu \nu} =
\begin{pmatrix}
-\alpha^2 + \beta_k \beta^k & \beta_j \\
\beta_j & \gamma_{ij}
\end{pmatrix},
\qquad
g^{\mu \nu} =
\begin{pmatrix}
-1/\alpha^2 & \beta^j / \alpha^2 \\
\beta^j / \alpha^2 & \gamma^{ij} - \beta^i \beta^j / \alpha^2
\end{pmatrix},
\end{equation}

\noindent and the spatial projection operator and the metric corresponding to each slice, $\Sigma_t$, can be expressed as: 

\begin{equation}
\gamma_{\mu \nu} := g_{\mu \nu} + n_\mu n_\nu, \qquad \gamma^\mu{}_\nu := \delta^\mu{}_\nu + n^\mu n_\nu,\label{eq:metric_and_inverse}
\end{equation}

\noindent where $\gamma_{ij}$ is the spatial part of $g_{\mu\nu}$, and $\gamma^{ij}$ is the algebraic inverse of $\gamma_{ij}$. The shift vector components satisfy $\beta^{i} = \gamma^{ij}\beta_{j}$, and the determinants of the metrics are related by $(-g) = \alpha^{2}\gamma$. Using Eq.~\eqref{eq:metric_and_inverse}, we can project any four-vectors or tensors to their temporal and spatial components. 
To define the fluid three-velocity, we use the four-velocity of a fluid element, $u^{\mu}$, which has a Lorentz factor $W := - u^{\mu} n_{\mu} = \alpha u^{0} = (1 - v^2)^{-1/2}$ with $v^2 := v_i v^i$:

\begin{equation}
\begin{aligned}
v^i := \frac{\gamma^i_\mu u^\mu}{W} = \frac{u^i}{W} + \frac{\beta^i}{\alpha}, \qquad
v_i := \gamma_{ij} v^j = \frac{u_i}{W}.
\end{aligned}
\end{equation}

\subsection{Maxwell's Equations}

Based on Maxwell’s equations, the extended electromagnetic field dynamics take the following form~\cite{palenzuela2010understanding}:

\vspace{-0.1em} 
\begin{align}
\nabla_\nu \left( F^{\mu\nu} + g^{\mu\nu} \psi \right) 
&= I^\mu - \kappa n^\mu \psi, 
\label{eq:maxwell1} \\[1ex]
\nabla_\nu \left( {}^*F^{\mu\nu} + g^{\mu\nu} \phi \right) 
&= -\kappa n^\mu \phi,
\label{eq:maxwell2}
\end{align}

\noindent where \( F^{\mu \nu} \) is the Faraday tensor, \( {}^*F^{\mu \nu} \) is its Hodge dual, and \( I^{\mu} \) is the electric current. 
We introduce the auxiliary scalar variables \( \psi \) and \( \phi \) into Maxwell’s equations to control the constraints associated with the magnetic and electric components. 
The Faraday tensor can be expressed in terms of its Hodge dual using the four-dimensional Levi-Civita tensor, $\epsilon^{\mu \nu \alpha \beta} \equiv \eta^{\mu \nu \alpha \beta} / \sqrt{-g}$ and $\epsilon_{\mu \nu \alpha \beta} \equiv \sqrt{-g} \, \eta_{\mu \nu \alpha \beta}$:

\begin{equation}
{}^{*}F^{\mu \nu} = \frac{1}{2} \epsilon^{\mu \nu \alpha \beta} F_{\alpha \beta}, \quad
F^{\mu \nu} = -\frac{1}{2} \epsilon^{\mu \nu \alpha \beta} {}^{*}F_{\alpha \beta}   .
\end{equation}

The Faraday tensor and its Hodge dual decompose into the electric and magnetic fields as measured by an observer traveling in the normal direction $n^\mu$ in the standard 3+1 splitting of spacetime:

\begin{align}
F^{\mu \nu} = n^\mu E^\nu - n^\nu E^\mu + \epsilon^{\mu \nu \alpha \beta} B_\alpha n_\beta, \\[1.5ex]
{}^*F^{\mu \nu} = n^\mu B^\nu - n^\nu B^\mu - \epsilon^{\mu \nu \alpha \beta} E_\alpha n_\beta,
\end{align} 

\noindent where the electric and magnetic field components are denoted by ${E}^\mu$ and ${B}^\mu$, respectively.
Similarly, we decompose the electric four-current $I^\mu$ as:

\begin{equation}
I^\mu := q n^\mu + J^\mu,
\end{equation}

\noindent where the charge density and the electric current density are represented by $q$ and $J^\mu$, respectively. The charge density $q$ is a scalar measured by the Eulerian observer~\cite{palenzuela2010understanding}.
Additionally, we express the electric and magnetic fields in the fluid framework (comoving frame) as follows~\cite{Bucciantini2012, Cheong2022}:

\begin{align}
e^\mu &:= F^{\mu\nu} u_\nu 
= W n^\mu (E^i v_i) + W \left( E^\mu + \epsilon^{\mu\nu\lambda} v_\nu B_\lambda \right), \\
b^\mu &:= {}^*F^{\mu\nu} u_\nu 
= W n^\mu (B^i v_i) + W \left( B^\mu - \epsilon^{\mu\nu\lambda} v_\nu E_\lambda \right),
\end{align}

\noindent then, we rewrite the electromagnetic energy–momentum tensor $T^{\mu \nu}_{EM}$ as:

\begin{equation}
T^{\mu\nu}_{\text{EM}} = (b^2 + e^2) \left( u^\mu u^\nu + \frac{1}{2} g^{\mu\nu} \right)
- b^\mu b^\nu - e^\mu e^\nu 
- \alpha u_\lambda e_\beta b_\kappa \left( u^\mu \epsilon^{\nu\lambda\beta\kappa} + u^\nu \epsilon^{\mu\lambda\beta\kappa} \right).
\end{equation}

\noindent
Here, \( \epsilon^{\mu \nu \lambda} \) is related to the four-dimensional Levi-Civita tensor by 
\( \epsilon^{\mu \nu \lambda} := n_\rho \epsilon^{\rho \mu \nu \lambda} \), 
and the comoving electric and magnetic field strengths are given by 
\( e^2 = e^\mu e_\mu \) and \( b^2 = b^\mu b_\mu \), respectively.
Using the 3+1 decomposition, the covariant Maxwell's equations \eqref{eq:maxwell1} and \eqref{eq:maxwell2} take the following form in terms of the divergence-cleaning scalars and electromagnetic fields~\cite{dionysopoulou2015general}:

\begin{align}
&\left( \partial_t - \mathcal{L}_{\beta} \right) E^i 
- \epsilon^{ijk} \nabla_j \left( \alpha B_k \right) 
+ \alpha \gamma^{ij} \nabla_j \psi 
= \alpha K E^i - \alpha J^i, \\[1.5ex]
&\left( \partial_t - \mathcal{L}_{\beta} \right) \psi 
+ \alpha \nabla_i E^i 
= \alpha q - \alpha \kappa \psi, \label{eq:Psi} \\[1.5ex]
&\left( \partial_t - \mathcal{L}_{\beta} \right) B^i 
+ \epsilon^{ijk} \nabla_j \left( \alpha E_k \right) 
+ \alpha \gamma^{ij} \nabla_j \phi 
= \alpha K B^i, \\[1.5ex]
&\left( \partial_t - \mathcal{L}_{\beta} \right) \phi 
+ \alpha \nabla_i B^i 
= -\alpha \kappa \phi, \label{eq:Phi}\\[1.5ex]
&\left( \partial_t - \mathcal{L}_{\beta} \right) q 
+ \nabla_i \left( \alpha J^i \right) 
= \alpha K q. \label{eq:charge_density}
\end{align}

\noindent Here, $\mathcal{L}_{\beta}$ denotes the Lie derivative along $\beta^i$. Note that the action of \( \mathcal{L}_{\beta} \) depends on the nature of the quantity: for a scalar \( f \), \( \mathcal{L}_{\beta} f = \beta^j \nabla_j f \); and for a vector \( V^i \), \( \mathcal{L}_{\beta} V^i = \beta^j \nabla_j V^i - V^j \nabla_j \beta^i \).
The quantity $K := K^i_{i} = \gamma^{ij} K_{ij}$ is the trace of the extrinsic curvature, and $\epsilon^{ijk} = \eta^{ijk} / \sqrt{\gamma}$ is the three-dimensional Levi-Civita tensor. The scalar fields $\psi$ and $\phi$, which measure the deviation from the constrained solution, respectively, push the solution of Eq.~\eqref{eq:Psi} towards the condition $\nabla_i E^i = q$ and the solution of Eq.~\eqref{eq:Phi} towards the zero-divergence condition $\nabla_i B^i = 0$.  This driving occurs over a timescale of $1/\kappa$ and at exponential speed~\cite{dionysopoulou2015general}. This approach, first presented in the context of IMHD as hyperbolic divergence cleaning, provides a simple way to solve Maxwell’s equations and enforce the divergence-free condition for the magnetic field. A resistive extension of this method was proposed by~\cite{Komissarov2007, Palenzuela2009}. 
We obtain the evolution equation for the charge density \eqref{eq:charge_density} from the conservation law associated with electric charge, $\nabla_\mu I^\mu = 0$.
We must add a prescription for the spatial components $J^i$, known as Ohm's law, in order to couple the EM fields with the fluid and complete the system of Maxwell's equations. We will present the full set of GR-RMHD evolution equations in the following section.

\subsection{Magnetohydrodynamics Equations} 

Together, the conservation of momentum and energy and the conservation of rest mass produce the fluid equations of general relativity:

\begin{equation}
\nabla_\mu T^{\mu \nu} = 0, \qquad \nabla_\mu (\rho u^\mu) = 0.
\end{equation}

\noindent We obtain the stress–energy tensor of a magnetized perfect fluid using~\cite{Palenzuela2013,Ripperda2019}:

\begin{equation}
T^{\mu \nu} = T^{\mu \nu}_{fluid} + T^{\mu \nu}_{EM} = \left[ \rho (1 + \epsilon) + p \right] u^\mu u^\nu + p g^{\mu \nu} + F^{\mu \alpha} F_\alpha{}^\nu - \frac{1}{4} g^{\mu \nu} F_{\alpha \beta} F^{\alpha \beta},
\end{equation}

\noindent where $\rho$ is the rest-mass density, $p$ and $\epsilon$ stand for fluid pressure and specific internal energy, respectively. Enthalpy is given by: $ h = \rho (1 + \epsilon ) + p$. 
We obtain the definition of the conserved variables from the projections of the stress–energy tensor:

\begin{align}
&D := \sqrt{\gamma} \rho W, \label{eq:conserved_dens}\\[1.5ex] 
&\tau := U - D = \sqrt{\gamma} \left(h W^2 - p + \frac{1}{2}(E^2 + B^2) \right) - D, \label{eq:conserved_tau} \\[1.5ex]
&S_i := \sqrt{\gamma} \left(hW^2 v_i + \epsilon_{ijk} E^j B^k \right). \label{eq:conserved_S}
\end{align}

\noindent The conserved rest mass density is represented by $D$, the energy density by $U$, and the conserved momentum by $S_i$. We denote Eq.~\eqref{eq:conserved_dens} to \eqref{eq:conserved_S} as conserved quantities because they follow from conservation laws in flat spacetime~\cite{Banyuls1997}. 
We also define the fully spatial projection of the energy–momentum tensor as:

\begin{align}
&S_{ij} := \sqrt{\gamma}\left(h W^2 v_i v_j + \gamma_{ij} p 
- E_i E_j - B_i B_j 
+ \frac{1}{2} \gamma_{ij} \left( E^2 + B^2 \right) \right).
\label{eq:conserved_Sij}
\end{align}

\subsection{Resistive GRMHD System}  

Whereas fluid fields are controlled by the conservation of total energy, momentum, and baryon number, electromagnetic fields evolve according to Maxwell's equations and charge conservation. 
Then, we solve these nonlinear equations in the presence of shocks by combining the MHD and Maxwell's equations in flux conservative form:

\begin{equation}
\partial_t \bm{U} + \partial_i \bm{F}^i = \bm{S},
\end{equation}

\noindent where $\bm{U}$ stands for the vector of conserved variables, $\bm{F}^i$ for the fluxes, and $\bm{S}$ for the source terms. Then, the evolution equations can be represented as~\cite{Palenzuela2013}: 

\begin{equation}
\label{eq:GRRMHD_flux_form}
\mathbf{U} =
\begin{bmatrix}
D \\[1ex]
S_j \\[1ex]
\tau \\[1ex]
\mathcal{B}^i \\[1ex]
\mathcal{E}^i \\[1ex]
\mathcal{\Phi} \\[1ex]
\mathcal{\Psi} \\[1ex]
\mathcal{Q}
\end{bmatrix},
\qquad
\mathbf{F}^i =
\begin{bmatrix}
(\alpha v^k - \beta^k) D \\[1ex]
\alpha\sqrt{\gamma}S^k_i - \beta^k S_i \\[1ex]
\alpha (S^k - v^k D) - \beta^k \tau \\[1ex]
\alpha\left(\epsilon^{ikj}\mathcal{E}_j + \gamma^{ik} \Phi \right) - \beta^k \mathcal{B}^i \\[1ex]
- \left[\alpha\left(\epsilon^{ikj}\mathcal{B}_j - \gamma^{ik} \Psi \right) + \beta^k \mathcal{E}^i \right] \\[1ex]
\alpha \mathcal{B}^k - \beta^k \Phi \\[1ex]
\alpha \mathcal{E}^k - \beta^k \Psi \\[1ex]
\alpha J^k - \beta^k \mathcal{Q}
\end{bmatrix},
\qquad
\mathbf{S} =
\begin{bmatrix}
0 \\[1ex]
\frac{\alpha \sqrt{\gamma}}{2} S^{jk} (\partial_i \gamma_{jk}) 
+ S_j (\partial_i \beta^j) - (\tau + D)(\partial_i \alpha) \\[1ex]
\alpha\sqrt{\gamma} S^{ij} K_{ij} - S^j (\partial_j \alpha) \\[1ex]
- \mathcal{B}^k (\partial_k \beta^i) + \Phi \left( \gamma^{ij} \partial_j \alpha - \alpha \gamma^{jk} \Gamma^i_{jk} \right) \\[1ex]
- \mathcal{E}^k (\partial_k \beta^i) + \Psi \left( \gamma^{ij} \partial_j \alpha - \alpha \gamma^{jk} \Gamma^i_{jk} \right) - \alpha J^i \\[1ex]
- \alpha \Phi \, K + \mathcal{B}^k (\partial_k \alpha) - \alpha \kappa \Phi \\[1ex]
- \alpha \Psi \, K + \mathcal{E}^k (\partial_k \alpha) + \alpha \mathcal{Q} - \alpha \kappa \Psi \\[1ex]
0
\end{bmatrix}.
\end{equation}

\noindent This structure yields a standard form of the equations, making them ideal for advanced numerical methods, especially those intended to solve hyperbolic partial differential equations, like high-resolution shock-capturing schemes. We express the conserved variables of the charge density, electric and magnetic fields, and the divergence-cleaning scalars as:

\begin{align}
\mathcal{E}^i = \sqrt{\gamma} E^i, \quad
\mathcal{B}^i = \sqrt{\gamma} B^i, \quad
\Psi^i = \sqrt{\gamma} \psi^i, \quad
\Phi^i = \sqrt{\gamma} \phi^i, \quad
\mathcal{Q} = \sqrt{\gamma} q.
\end{align}

Closing the system of equations in the GR-RMHD framework requires a generalized version of Ohm's law that permits the relationship between the electric current density, the electromagnetic fields, and the fluid's properties. The electric current can be modeled in the relativistic framework using a combination of an advective and a conductivity component.
When the collision frequency of particles is significantly higher than the characteristic frequency of plasma oscillations, the electrical conductivity tensor can be presumed to be isotropic~\cite{dionysopoulou2015general}. This simplification reduces the four-current equation to a closed form in which the spatial component of the current density in the Eulerian system is defined as a function of the electric field and fluid velocity:
\begin{equation}
J^i = \mathcal{Q}v^i + W \sigma \left[ \mathcal{E}^i + \epsilon^{ijk} v_j \mathcal{B}_k - \left( v_k \mathcal{E}^k \right) v^i \right], 
\label{eq:Ohm_law}
\end{equation}

\noindent where $\sigma$ is the electrical conductivity and is inversely proportional to the resistivity, i.e. $\sigma = 1/\eta$. The design of this formulation allows it to naturally recover the ideal MHD equations in the very high conductivity limit while also accounting for non-ideal phenomena like the decoupling between the fluid and the magnetic field in low-density regions.

\section{Numerical methods}
\label{sec:Numerical_methods}

\texttt{GRaM-X} (\textbf{G}eneral \textbf{R}elativistic \textbf{a}ccelerated-\textbf{M}agnetohydrodynamics on \textbf{A}MReX) is a GPU-accelerated GRMHD code that has originally been implemented for the equations of ideal GRMHD~\cite{Shankar2023}. \texttt{GRaM-X} uses a new adaptive mesh refinement~(AMR) driver for the Einstein Toolkit~\cite{Lffler2012, Msta2013, ZILHO2013} called \texttt{CarpetX} \cite{https://doi.org/10.5281/zenodo.6131529, Shankar2023, Kalinani2024} to enable AMR on GPUs. The code uses the Z4c~\cite{Ruiz2011, Hilditch2013} formalism to evolve the equations of general relativity. \texttt{GRaM-X} incorporates the HLLE Riemann solver, as well as the TVD~\cite{Toro2009} and WENO~\cite{Shu1998} reconstruction methods. 
\texttt{GRaM-X}/\texttt{\texttt{CarpetX}} can in principle utilize three levels of parallelism to optimize the use of modern computational resources: 1) shared memory parallelism (multi-threading), 2) accelerator-based parallelism (GPU), and 3) distributed memory parallelism (MPI). The entire computational domain is  divided into "boxes" which are distributed to various MPI ranks across compute nodes. When running on CPUs, each "box" is further divided into smaller logical tiles, with each tile executed by its own OpenMP thread. When running on GPUs, each cell in the "box" is executed by individual GPU threads because, unlike CPUs, a single GPU can execute thousands of threads at once. Halo data between "boxes" is synchronized using ghost zones, providing consistency throughout the entire simulation domain. For this new GR-RMHD module and in general for \texttt{GRaM-X}~\cite{Shankar2023}, we focus on 2) and 3) to utilize modern GPU supercomputers.

In this work, we implement a new resistive GRMHD module within \texttt{GRaM-X} i.e., we extend the GRMHD equations to the GR-RMHD equations. The equations of GR-RMHD are stiff due to the presence of conductivity $\sigma$, which can vary over orders of magnitude spatially and temporally. In order to evolve this system of equations stably, we need to use either explicit methods with very small timesteps or implicit methods, both of which are very expensive. In this work, we implement an Implicit-Explicit scheme that treats stiff terms implicitly and non-stiff terms explicitly, enabling stable evolution of the GR-RMHD equations without the need for fully implicit solvers or prohibitively small timesteps. We use the same reconstruction methods that are already implemented for ideal GRMHD within \texttt{GRaM-X}. However, we cannot use the same Riemann solver because the characteristic speeds for the GR-RMHD equations are different. For simplicity, we use the characteristic speed required in the Riemann solver as the speed of light, in which case the characteristic speed $\lambda_{\pm}^i$ in the $i$-th direction becomes  

\begin{align}
\lambda_{\pm}^i = \pm\sqrt{\gamma^{ii}} - \frac{\beta^i} {\alpha},
\end{align}
where $\gamma^{ii}$ is the diagonal component of the 3-metric.

\subsection{Implicit-Explicit Runge-Kutta Methods}

Although the ideal GRMHD equations are considered highly efficient for numerical implementation due to their relatively simple structure and favorable computational properties, the full formulation of GR-RMHD in a relativistic framework becomes significantly more complex when the electrical conductivity of the plasma exhibits strong spatial variations.
In these circumstances, the system's temporal evolution can vary significantly between areas with different conductivities; regions with high and low conductivities evolve on different and frequently incompatible timescales.

In the case of GR-RMHD equations, we have a hyperbolic system with stiff relaxation terms, in which some of the equations' components change over significantly shorter timescales than others. This stiffness reduces the numerical stability of simulations and necessitates the use of specialized algorithms that offer high accuracy and robust control to ensure accurate modeling of system dynamics. A system of partial differential equations with stiff terms can be written as follows~\cite{Palenzuela2009}:

\begin{equation}
\partial_t \bm{U} = F(\bm{U}) + \frac{1}{\epsilon} R(\bm{U}),
\label{eq:IMEX_eq}
\end{equation}

\noindent where $F(\bm{U})$ represents the non-stiff (explicit) terms of the GR-RMHD system and $R(\bm{U})$ contains the stiff relaxation terms scaled by $\epsilon$.  $F(\bm{U})$ should not be confused with the flux function $F^i(\bm{U})$ introduced earlier.  In this model, $\epsilon > 0$ represents the relaxation time, which may vary in space or time. In the limit $\epsilon \to \infty$, the equations reduce to a hyperbolic system with a finite characteristic wave speed $c_h$, defining a hyperbolic timescale $\tau_h = L / c_h$, where $L$ is a characteristic length scale. As $\epsilon$ decreases, approaching the infinite conductivity limit, the system becomes more and more stiff, and the relaxation timescale becomes much shorter than $\tau_h$.
In such a regime, explicit numerical methods remain stable only if the timestep satisfies $\Delta t \lesssim \epsilon$, which is significantly more restrictive than the conventional CFL condition. This imposes serious challenges for numerical integration, particularly when $\epsilon$ varies widely across the computational domain and is much smaller than $\tau_h$. As a result, designing robust and efficient numerical schemes for such systems becomes a major computational and algorithmic challenge.

To address the challenges posed by temporal stiffness in solving the resistive GRMHD equations within a uniform computational framework, we use an approach based on implicit–explicit Runge-Kutta schemes. This method is specifically designed to ensure numerical stability in the presence of stiff source terms and to enable accurate resolution of multiscale dynamical behavior~\cite{50c40ae4-606f-3fe2-a754-5dcba35beb0f, ASCHER1997151,Pareschi2001}. One of the key advantages of the IMEX scheme over a fully implicit method is that its implicit part can be solved point-wise at each grid cell. This approach avoids solving a large global linear system that couples all grid points simultaneously, which would be computationally prohibitive. Instead, the IMEX scheme solves many small systems independently, making the method significantly more efficient and easily parallelizable~\cite{Pareschi2005}.
IMEX methods work by combining two different discretization strategies: they treat the stiff terms using an implicit scheme and handle the non-stiff parts with an explicit one. When this approach is applied to the system of equations \eqref{eq:IMEX_eq}, it leads to the following form~\cite{Pareschi2005}:

\begin{equation}
\begin{aligned}
\bm{U}^{(i)} &= \bm{U}^n + \Delta t \sum_{j=1}^{i-1} \tilde{a}_{ij} F\left[\bm{U}^{(j)}\right] 
+ \Delta t \sum_{j=1}^{N} a_{ij} \frac{1}{\epsilon} R\left[\bm{U}^{(j)}\right], \\
\bm{U}^{n+1} &= \bm{U}^n + \Delta t \sum_{i=1}^{N} \tilde{\omega}_i F\left[\bm{U}^{(i)}\right] 
+ \Delta t \sum_{i=1}^{N} \omega_i \frac{1}{\epsilon} R\left[\bm{U}^{(i)}\right].
\end{aligned}
\end{equation}

\noindent In this formulation, \( \bm{U}^{(i)} \) denotes the intermediate stage values and \( \bm{U}^{n+1} \) represents the updated solution.  
The scheme consists of \( N \) Runge–Kutta stages, with \( \tilde{A} = (\tilde{a}_{ij}) \) and \( A = (a_{ij}) \) denoting the explicit and implicit Butcher matrices, respectively, of size \( N \times N \), and \( \tilde{\boldsymbol{\omega}} = (\tilde{\omega}_i) \), \( \boldsymbol{\omega} = (\omega_i) \) their corresponding weight vectors.  
The coefficients are chosen such that \( \tilde{a}_{ij} = 0 \) for \( j \ge i \) (explicit part) and \( a_{ij} = 0 \) for \( j > i \) (implicit diagonal structure), the latter corresponding to an L-stable Diagonally Implicit Runge–Kutta (DIRK) scheme that simplifies the implicit solve at each stage.
A standard approach for representing the structure of such methods is the Butcher notation, where the full specification of the scheme is expressed through a double Butcher tableau~\cite{butcher1987}

\begin{equation}
\begin{array}{c|c}
\tilde{c} & \tilde{A} \\
\hline
         & \tilde{\omega}^{T}
\end{array}
\qquad
\begin{array}{c|c}
c & A \\
\hline
  & \omega^{T}
\end{array}
\end{equation}

\noindent where superscript index $T$ denotes the transpose operator, and the coefficient vectors \( \tilde{c}\) and $c$, which are used in the analysis of non-autonomous systems, are defined as follows:

\begin{equation}
\tilde{c}_i = \sum_{j=1}^{i-1} \tilde{a}_{ij}, \qquad
c_i = \sum_{j=1}^{i} a_{ij}.
\end{equation}

We implement the IMEX-RK SSP2 $(2, 2, 2)$ L-stable scheme represented by the coefficient tableau in Table~\ref{table:imex}, with its intermediate and final steps given explicitly as follows: 

\begin{align}
\bm{U}^{(1)} &= \bm{U}^n + \frac{\Delta t}{\epsilon} \gamma R[\bm{U}^{(1)}], \\
\bm{U}^{(2)} &= \bm{U}^n + \Delta t F[\bm{U}^{(1)}] 
+ \frac{\Delta t}{\epsilon} \left\{ (1 - 2\gamma) R[\bm{U}^{(1)}] + \gamma R[\bm{U}^{(2)}] \right\}, \\
\bm{U}^{n+1} &= \bm{U}^n + \frac{\Delta t}{2} \left[ F(\bm{U}^{(1)}) + F(\bm{U}^{(2)}) \right] 
+ \frac{\Delta t}{2\epsilon} \left\{ R[\bm{U}^{(1)}] + R[\bm{U}^{(2)}] \right\}. \label{eq:IMEX3} 
\end{align}

\begin{table}[h!]
\centering
\[
\begin{array}{c|cc}
0 & 0 & 0 \\
1 & 1 & 0 \\
\hline
  & 1/2 & 1/2
\end{array}
\qquad\quad
\begin{array}{c|cc}
\gamma & \gamma & 0 \\
1-\gamma & 1 - 2\gamma & \gamma \\
\hline
  & 1/2 & 1/2
\end{array}
\]
\[
\gamma \equiv 1 - \frac{1}{\sqrt{2}}.
\]
\caption{Butcher tableaux for the IMEX-RK SSP2 $(2, 2, 2)$ L-stable scheme. Left: explicit tableau; Right: implicit tableau.}
\label{table:imex}
\end{table}
\begin{center}
\rule{10cm}{0.4pt}
\end{center}

In the system of Eq.~\eqref{eq:GRRMHD_flux_form}, we can separate the evolving fields into two categories: the \textit{stiff} ones, which are the electric field components grouped as \( X = \{ \mathcal{E}^i_\mathrm{stiff} \} \), and the \textit{non-stiff} ones, which cover the rest of the dynamic variables grouped as \( Y = \{\mathcal{E}^i_\mathrm{non-stiff}, \mathcal{B}^i, \Psi, \Phi, \mathcal{Q}, \tau, S_i, D \} \). In this setup, the equation governing the evolution of the electric field can become stiff when the electrical conductivity is high, and this situation is described by Ohm’s law (Eq.~\eqref{eq:Ohm_law}). We naturally split the right-hand side of the electric field equation into two parts: one that may cause stiffness in the system and another that behaves more smoothly and does not require special treatment:

\begin{equation}
\partial_t \mathcal{E}^i = F_E^i + R_E^i,
\end{equation}

\noindent where $F_E^i$ and $R_E^i$ are defined as follows:

\begin{align}
F_E^i &= - \partial_k \left[ -\beta^k \mathcal{E}^i  - \alpha \left( \epsilon^{ikj} \mathcal{B}_j - \gamma^{ik} \Psi \right) \right]
- \mathcal{E}^k (\partial_k \beta^i) \\
&\quad + \Psi \left( \gamma^{ij} \partial_j \alpha - \alpha \gamma^{jk} \Gamma^i_{jk} \right)
- \alpha \mathcal{Q} v^i, \\[1ex]
R_E^i &= -\alpha W \sigma \left[ \mathcal{E}^i + \epsilon^{ijk} v_j \mathcal{B}_k - (v_k \mathcal{E}^k) v^i \right].
\end{align}

Due to its linear dependence on the electric field, we handle the implicit stage involving the term $R_E^i(\mathcal{E}^{(i)})$, derived from the resistive Ohm’s law, either analytically or with a three-dimensional Newton-Raphson (3D-NR) solver.
In this work, we adopt the latter approach as our numerical solution strategy.
We determine the evolution of stiff electric field components using the roots of the following function:

\begin{align}
f(\mathcal{E}^i) &= \mathcal{E}^i - \mathcal{E}^i_{\text{initial}} + \Delta t \Big[ 
    \alpha W \sigma a_{ii}\left( \mathcal{E}^i 
    + \epsilon^{ijk} v_j \mathcal{B}_k - (v_k \mathcal{E}^k)v^i\right)     
\Big], \label{eq:3D_NR}
\end{align}

\noindent where $\mathcal{E}^i$ denotes the updated components of the electric field solved at the IMEX-RK2 current stage, while $\mathcal{E}^i_{\text{initial}}$ represents the initial values at the beginning of the implicit substeps. The term inside the square brackets corresponds to the stiff relaxation term from Ohm’s law, which is treated implicitly within the IMEX-RK2 update.
We calculate the solution at iteration $n+1$ as

\begin{equation}
{\mathcal{E}^i}^{(n+1)} = {\mathcal{E}^i}^{(n)} - \frac{f({\mathcal{E}^i}^{(n)})}{f'({\mathcal{E}^i}^{(n)})}
\qquad\qquad
\text{where} \quad\qquad 
f'({\mathcal{E}^i}^{(n)}) = \frac{df({\mathcal{E}^i}^{(n)})}{d{\mathcal{E}^i}^{(n)}}.
\end{equation}

It is important to note that updating the electric field requires knowledge of the three components of velocity \( v^i \) at the same time as \( \mathcal{E}^i \). However, since \( v^i \) is a primitive variable that depends nonlinearly on the electric field, this introduces an implicit coupling into the update equations. This recovery step is a critical part of any GR-RMHD simulation. It is highly sensitive to the underlying physical parameters and, due to the nonlinear nature of the system, can become one of the most technically demanding components of the numerical algorithm. Therefore, we must evolve $\mathcal{E}^i$ within the IMEX framework together with recovering the primitive variables from the conserved ones.

\subsection{Recovery of Primitive Variables from Conserved Variables}

The primitive variables ($\rho$, $\epsilon$, $p$, $v^i$, $E^i$, $B^i$, $\psi$, $\phi$, $q$) must be recovered from the evolved conserved variables ($D$, $\tau$, $S_i$, $\mathcal{E}^i$, $\mathcal{B}^i$, $\Psi$, $\Phi$, $\mathcal{Q}$) following each timestep in order to numerically solve the GR-RMHD equations. Despite the fact that the conserved fields are algebraic functions of the primitive ones, dependencies on the Lorentz factor and enthalpy make the inverse relation difficult to understand. With the exception of very basic equations of state (EoS), this makes the recovery process non-trivial and typically necessitates numerical root-finding.
We can compute most conserved quantities directly at time $t=(n+1)\Delta t$ by evolving their non-stiff evolution equations. However, for stiff components such as the electric field, the explicit update only offers a partial solution. To obtain a complete solution, we must solve an implicit equation with relaxation terms that rely on velocity and other primitive fields.

The recovery and implicit evolution steps are inherently coupled since the electric field, as mentioned before, depends on primitive variables, such as velocity, and the conserved ones depend on the electric field. Consequently, the complete solution entails solving both procedures at the same time, usually using an iterative algorithm that guarantees consistency between the evolution of the stiff terms and the primitive recovery. We use a one-dimensional approach to recover the primitive variables, following~\cite{Palenzuela2009}, and summarize the recovery process as follows:

(i) We use the values from the previous timestep to guess the pressure $p = p^n$ and $v^i = {v^{i}}^{(n)}$.

(ii) We use an iterative 3D-NR solver, as described in the previous section, to extract the components of $\mathcal{E}^i$ using the initial guess for the velocity and electric field $\mathcal{E}^i = {\mathcal{E}^i}^{(n)}$.

(iii) We compute the velocity, Lorentz factor, rest-mass density, and specific internal energy in the following order using the updated electric field value and the value of the pressure from a previous timestep:

\begin{align}
{v}_i &= \frac{S_i - \epsilon_{ijk}\mathcal{E}^j\mathcal{B}^k}{\tau + D +  \sqrt{\gamma} \ p - \frac{(\mathcal{E}^2 + \mathcal{B}^2)}{2 \sqrt{\gamma}}}, \\[1.5ex]
W &= \frac{1}{\sqrt{1 - v^iv_i}}, \\[1.5ex]
\rho &= \frac{D}{\sqrt{\gamma} \ W}, \\[1.5ex]
\epsilon &= \frac{\tau - D (W -1) - \sqrt{\gamma} \ p(W^2 -1 ) - \frac{(\mathcal{E}^2 + \mathcal{B}^2)}{2\sqrt{\gamma}}}{DW}.
\end{align}

\noindent Furthermore, we calculate the primitive variables for the electric field, magnetic field, scalar fields $\psi$ and $\phi$, and electric charge density by dividing the respective conserved variables by the square root of the determinant of the spatial 3-metric, $\sqrt{\gamma}$.

(iv) We numerically determine the pressure, similar to the electric field calculation, using an iterative Newton-Raphson solver. The solution at the next iteration is computed as follows:

\begin{align}
p^{n+1} = p^n - \frac{f(p^n)}{f'(p^n)},
\end{align}

\noindent where $f(p = {p}^n) = p(\rho(p^n), \ \epsilon(p^n)) - p^n$ and $f'(p = {p}^n) = v^2 c_s^2 - 1$~\cite{Noble2006}. $ p(\rho, \epsilon)$ is defined by the ideal-fluid EoS, and we determine the local speed of the fluid as follows: 

\begin{align}
&p(\rho, \epsilon) = (\Gamma - 1)\rho \epsilon, \\[1.5ex]
&c_s = \left( \frac{\Gamma (\Gamma - 1)}{1 + \Gamma \epsilon} \right)^{1/2}.
\end{align}

(v) After the initial pressure guess in step (i), we repeat steps (ii)–(iv) iteratively, updating the pressure and velocity until the change between successive iterations falls below a specified tolerance.

This method provides a straightforward and practical way to recover the primitive variables, especially when the magnetic pressure remains moderate compared to the fluid pressure (i.e. , ${|\mathcal{B}|^2}/p$). In such conditions, the algorithm typically converges in less than ten iterations and performs reliably for both continuous and discontinuous electromagnetic fields in the intermediate conductivity regions. However, $\lesssim 70$ iterations are typically needed to get convergence in high conductivity regions.

\section{Numerical Tests}
\label{sec:Numerical_Tests}

In this section, we present the results of tests we have conducted to validate the implementation of the IMEX-RK2 scheme across several regimes of GR-RMHD. We have adopted a hyperbolic divergence cleaning damping factor of $\kappa = 1$ for all the tests. The conducted tests include a one-dimensional self-similar current sheet, a shock-tube problem, large-amplitude CP Alfv\'{e}n waves, a two-dimensional cylindrical explosion, and the three-dimensional TOV star. In all these tests, we assume an ideal-fluid EoS, with $\Gamma = 2$ for all the one-dimensional and three-dimensional tests, and $\Gamma = 4/3$ for the two-dimensional cylindrical explosion test.

\subsection{1D MHD Shocktube}

\begin{figure}[H]
    \centering
    \includegraphics[height=5.9cm, width=7.9cm]{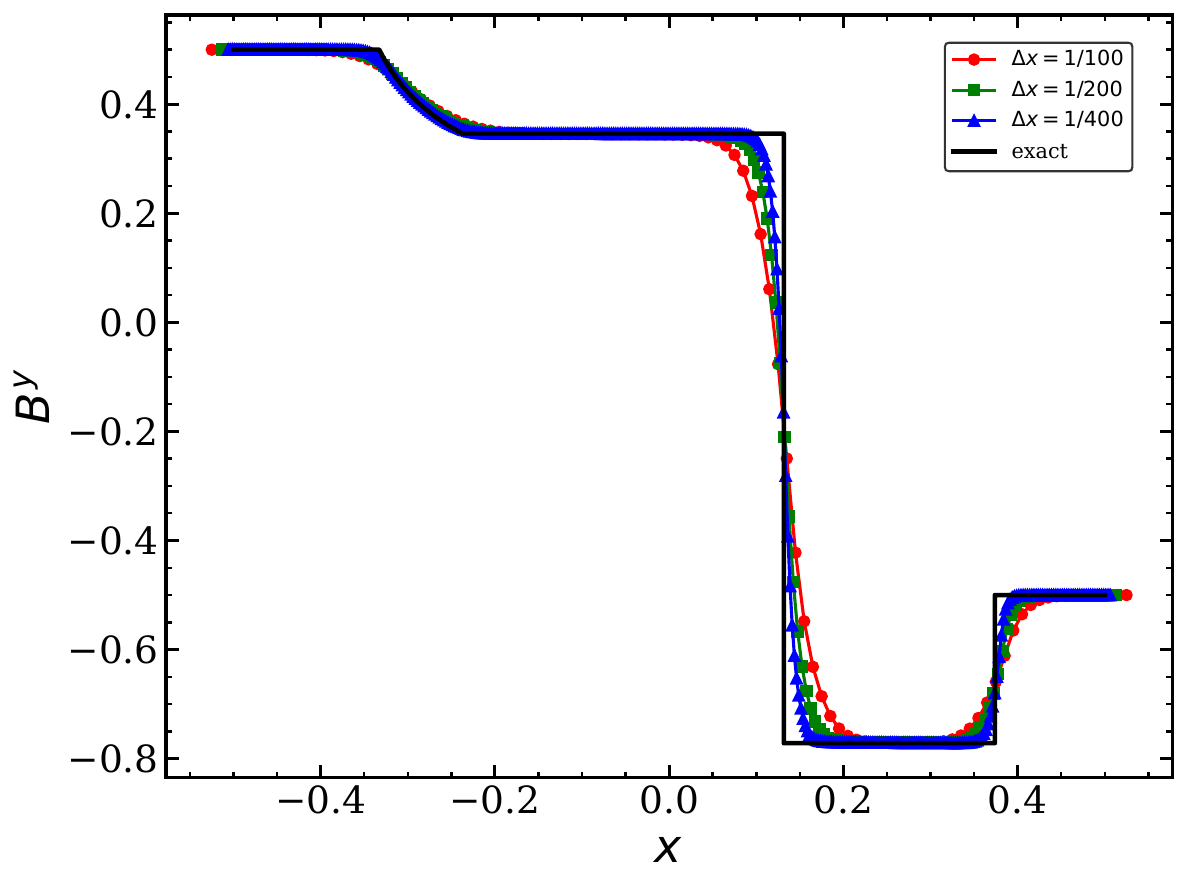}
    \hspace{0.1cm}
    \includegraphics[height=5.9cm, width=7.9cm]{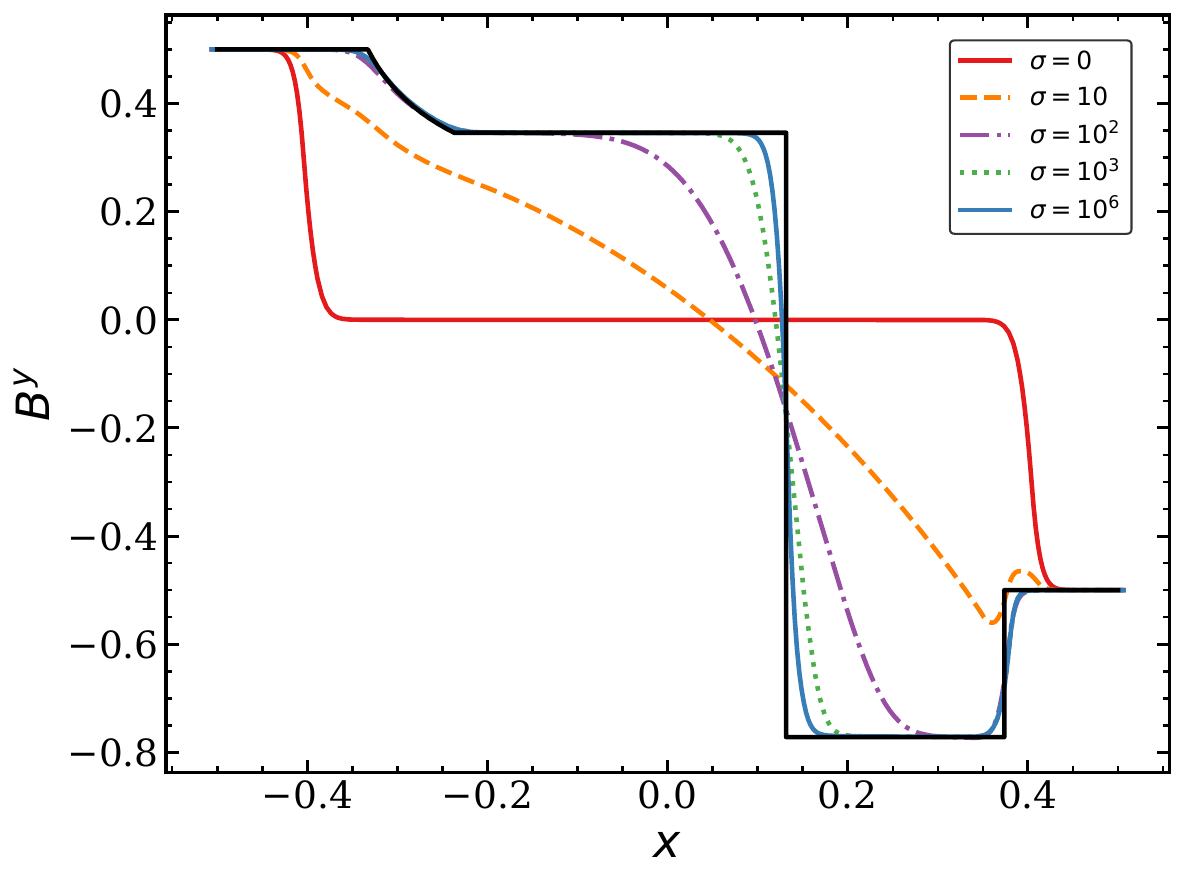}
    \caption{\small
        Left panel: Magnetic field component $B^y$ as a function of $x$ for the shock-tube problem. 
        The figure shows three lines, each corresponding to a different resolution at $t = 0.4$. The conductivity is uniform, with a value of $\sigma_0 = 10^6$. Right panel: Magnetic field component $B^y$ as a function of $x$ for varying uniform conductivities. As the conductivity rises, the solution approaches the ideal MHD solution. It is important to observe that when $\sigma_0 =0$, the solution represents a discontinuity traveling at the speed of light, which corresponds to Maxwell's equations in a vacuum.
    }
    \label{fig:const_shock}
\end{figure}

The planar magnetic shocktube tests are among the most effective ways to assess how well magnetohydrodynamics codes perform. We consider a shocktube test similar to~\cite{Palenzuela2009} to validate our code in various conductivity regimes, including the ideal MHD limit. This test was originally introduced by~\cite{Brio1988} and then a modified version of it was presented by~\cite{GIACOMAZZO2006}. In this magnetic shocktube test, $x=0$ acts as the plane of discontinuity for the Riemann problem, and the domain in the $x$-direction is divided into two parts, left $x<0$ and right $x>0$. The initial states on the left and right are defined as

\begin{equation}
(\rho, p, v^x, v^y, v^z, B^x, B^y, B^z) =
\begin{cases}
(1.0, \ 1.0, \ 0.0, \ 0.0, \ 0.0, \ 0.0, \ 0.5, \ 0.0), & x < 0 \\
(0.125, \ 0.1, \ 0.0, \ 0.0, \ 0.0, \ 0.0, -0.5, \ 0.0),         & x > 0
\end{cases}
\end{equation}
with the electric field set to zero. By setting $B^x = 0$, we obtain a structure consisting of two rapid waves: one propagating to the left (rarefaction) and the other to the right (shock), separated by a tangential discontinuity.
We have conducted this test in $x$-, $y$-, and $z$-directions, but only show the results in $x$-direction, noting that in the other directions the results show identical behavior. We use Neumann boundary conditions wherein the data is copied from the nearest interior point to the points on the boundary while the flux at the boundary points is set to zero. We reconstruct the primitive variables using TVD reconstruction with the minmod limiter.
We test both uniform and non-uniform conductivities, with the conductivity in the latter scenario being set to 

\begin{align}
\sigma = \sigma_0D^\gamma, \label{eq:sigma_evol}
\end{align}

\noindent where \( \sigma_0 \) and \( \gamma \) are positive constants (the latter not to be confused with the determinant of the spatial metric, \( \gamma = \det(\gamma_{ij}) \)).
Here, the conductivity has a nonlinear dependence on the conserved variable $D$, characterized by the exponent $\gamma$. We first consider the case with uniform conductivity, which we obtain by setting $\gamma$ equal to zero in Eq.~\eqref{eq:sigma_evol}. In this case, when the conductivity is very large ($\sigma_0 = 10^6$), the solution approaches the ideal MHD limit. We choose the test domain $[-0.5, 0.5]$, and employ resolutions $\Delta x = 0.01$, $0.005$, and $0.0025$,  with the CFL set to $0.25$. We show $B^y(x)$ for these three resolutions at time $t = 0.4$ in the left panel of Fig.~\ref{fig:const_shock} together with the exact solution in the ideal-MHD limit. The result demonstrates that, even in the presence of shocks, our resistive MHD numerical solution converges towards the ideal MHD solution as resolution increases. We plot $B^y(x)$ in the right panel of Fig.~\ref{fig:const_shock} for various uniform conductivities (i.e., $\gamma = 0$) at a fixed resolution $\Delta x = 0.0025$. Here we can see that the wave-like solution at $\sigma =\sigma_0 = 0$ gradually converges to the ideal MHD solution when we increase $\sigma_0$ up to $10^6$.
The solution of the shocktube in the presence of a non-uniform conductivity is one of the more challenging tests that we have carried out. In the non-uniform conductivity case, we vary $\gamma$ while keeping $\sigma_0$ fixed.

\begin{figure}[H]
    \centering
    \includegraphics[height=5.9cm, width=7.9cm]{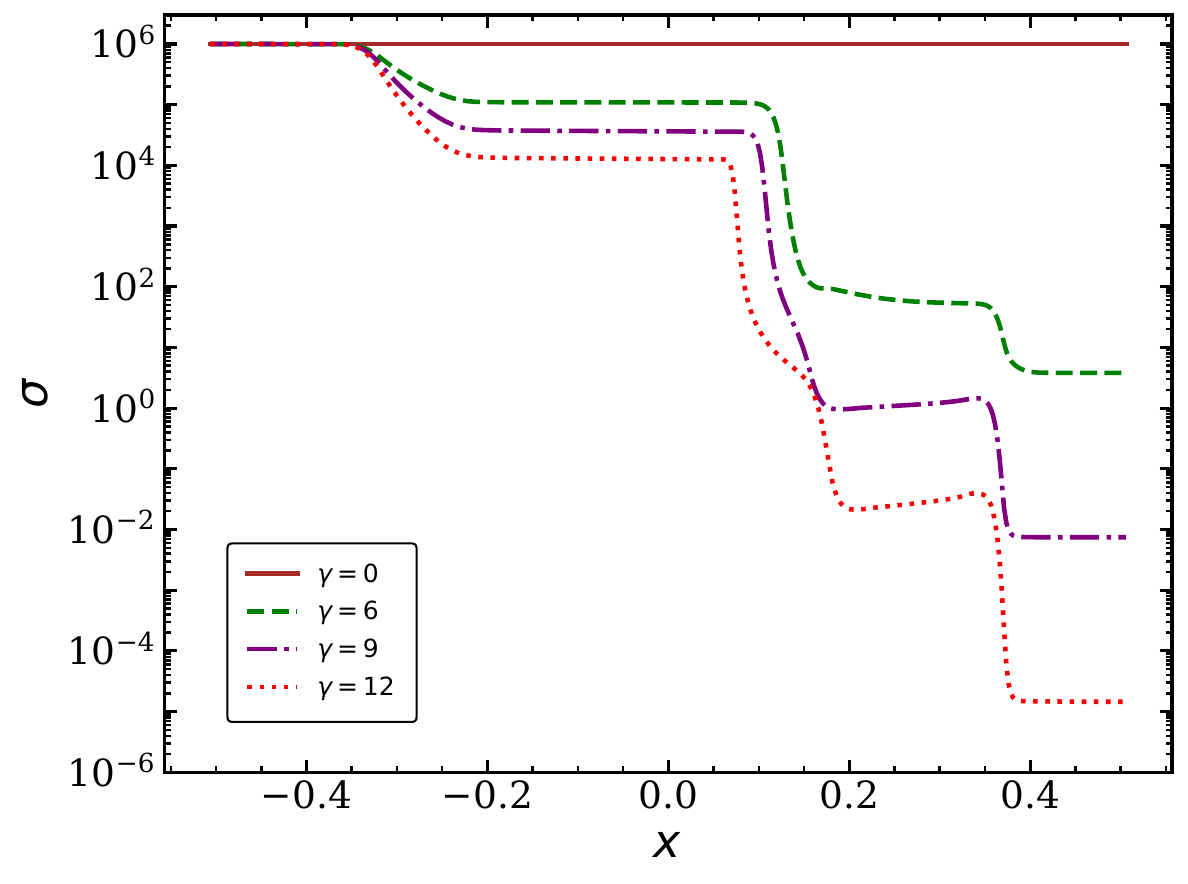}
    \hspace{0.1cm}
    \includegraphics[height=5.9cm, width=7.9cm]{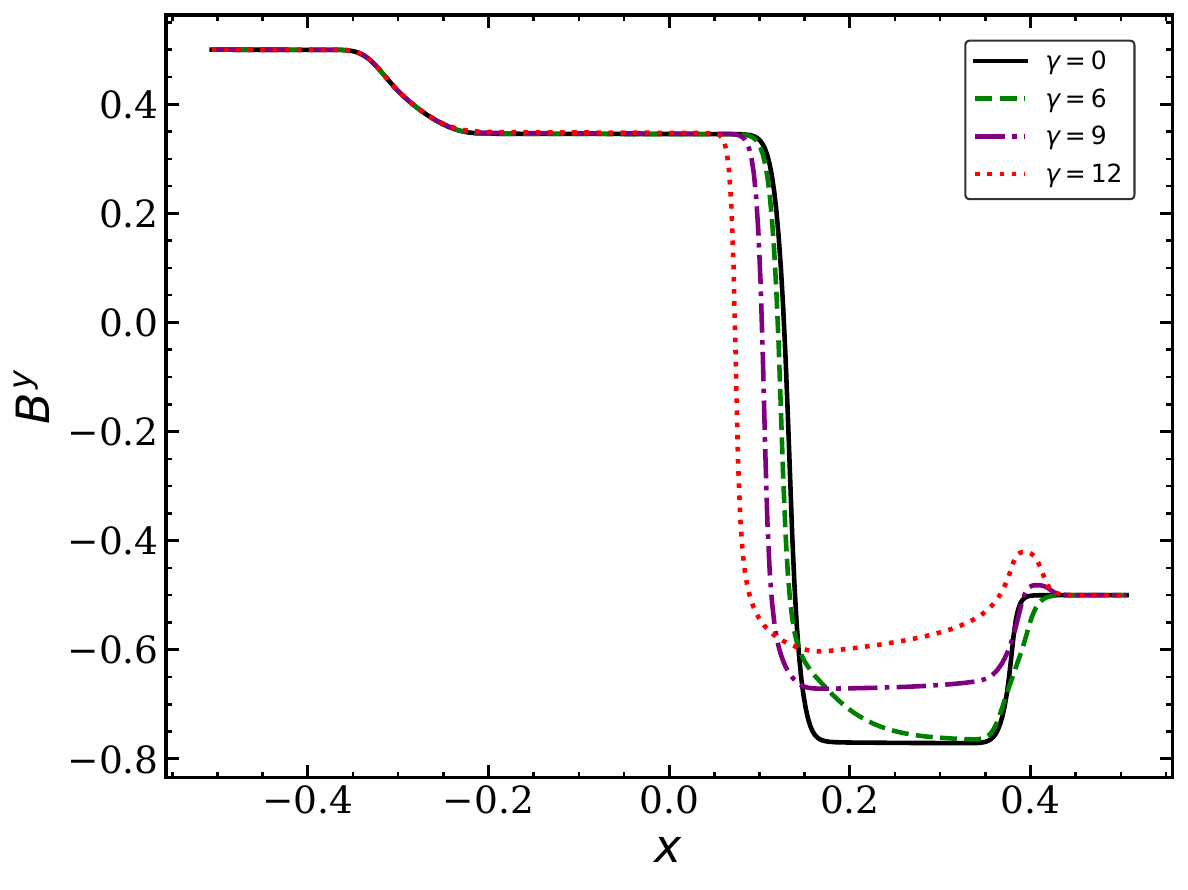}
     \caption{\small
        Left panel: The evolution of a non-uniform conductivity $\sigma$ in the shocktube problem for varying values of $\gamma$. Right panel: The magnetic field component $B^y$ as a function of $x$ for various values of $\gamma$ at $t = 0.4$. A fixed value of \( \sigma_0 = 10^6 \) is used in all cases.
    }
    \label{fig:dynamo_sigma}
\end{figure}

Next, we perform the shocktube tests for different choices of $\gamma = \{0, 6, 9,1 2\}$ at a constant $\sigma_0 = 10^6$ and $\Delta x = 0.0025$. We plot the conductivities at the final time $t=0.4$ in the left panel of Fig.~\ref{fig:dynamo_sigma}. In this particular scenario, the conductivity is set as a function of the rest-mass density, and our numerical scheme produces a stable and accurate solution even when the conductivity varies by nearly 12 orders of magnitude, ranging from $\sim 10^{-6}$ to $\sim 10^{6}$ depending on position across the computational grid.
Likewise, in the right panel, we show $B^y$ for various values of $\gamma$ at $t=0.4$.  
We can see in the left panel of Fig.~\ref{fig:dynamo_sigma} that the conductivity is very high ($\gtrsim10^4$) in the region $x<0$ for all values of $\gamma$, and hence the solution for $B^y$ in this region is expected to approach the ideal MHD limit. On the other hand, in the region $x>0$, where the rest-mass density is very low, the conductivity becomes small ($\lesssim1$) when $\gamma \gtrsim 6$, which means the solution for $B^y$ is expected to deviate from the ideal MHD limit. The results in this section are in good agreement with those of~\cite{Palenzuela2009, Dionysopoulou2013} and demonstrate that our code can effectively handle both uniform and non-uniform conductivity profiles.

\subsection{2D Cylindrical Explosion}
To further examine the capabilities of our code, we conduct a two-dimensional test, a cylindrical explosion expanding within a plasma permeated by an initially uniform magnetic field in the x-direction. 
This test provides a thorough assessment of the numerical implementation, since shocks develop and propagate in all directions within the two-dimensional domain. The cylindrical explosion test offers a straightforward setup that challenges the MHD scheme and reveals possible hidden flaws in the implementation. An exact solution for this test does not exist; however, we compare our results with those from other codes in the literature.
\begin{figure}[H]
    \centering
    \includegraphics[height=7.8cm, width=8.8cm]{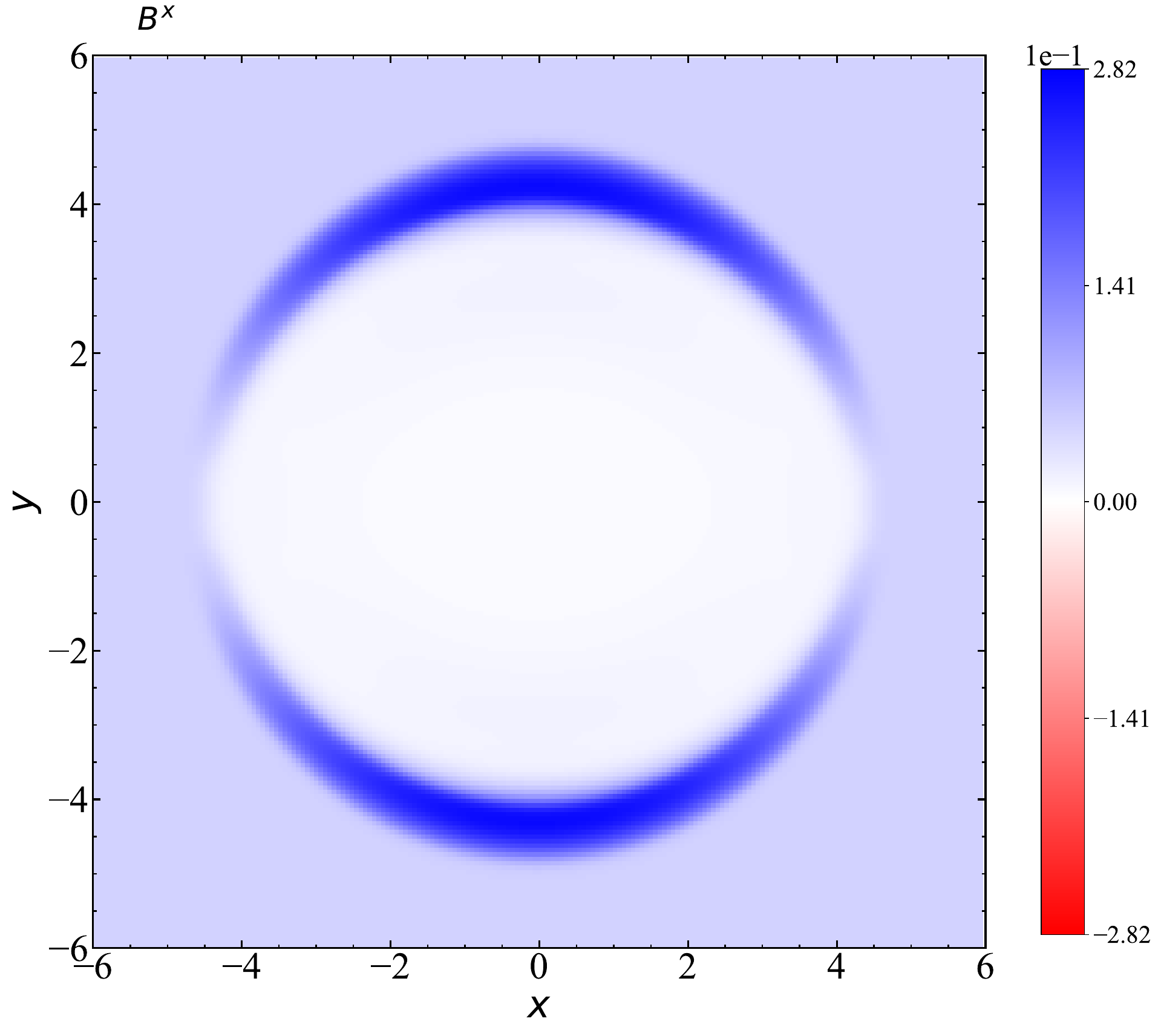}
    \hspace{0.1cm}
    \includegraphics[height=7.8cm, width=8.8cm]{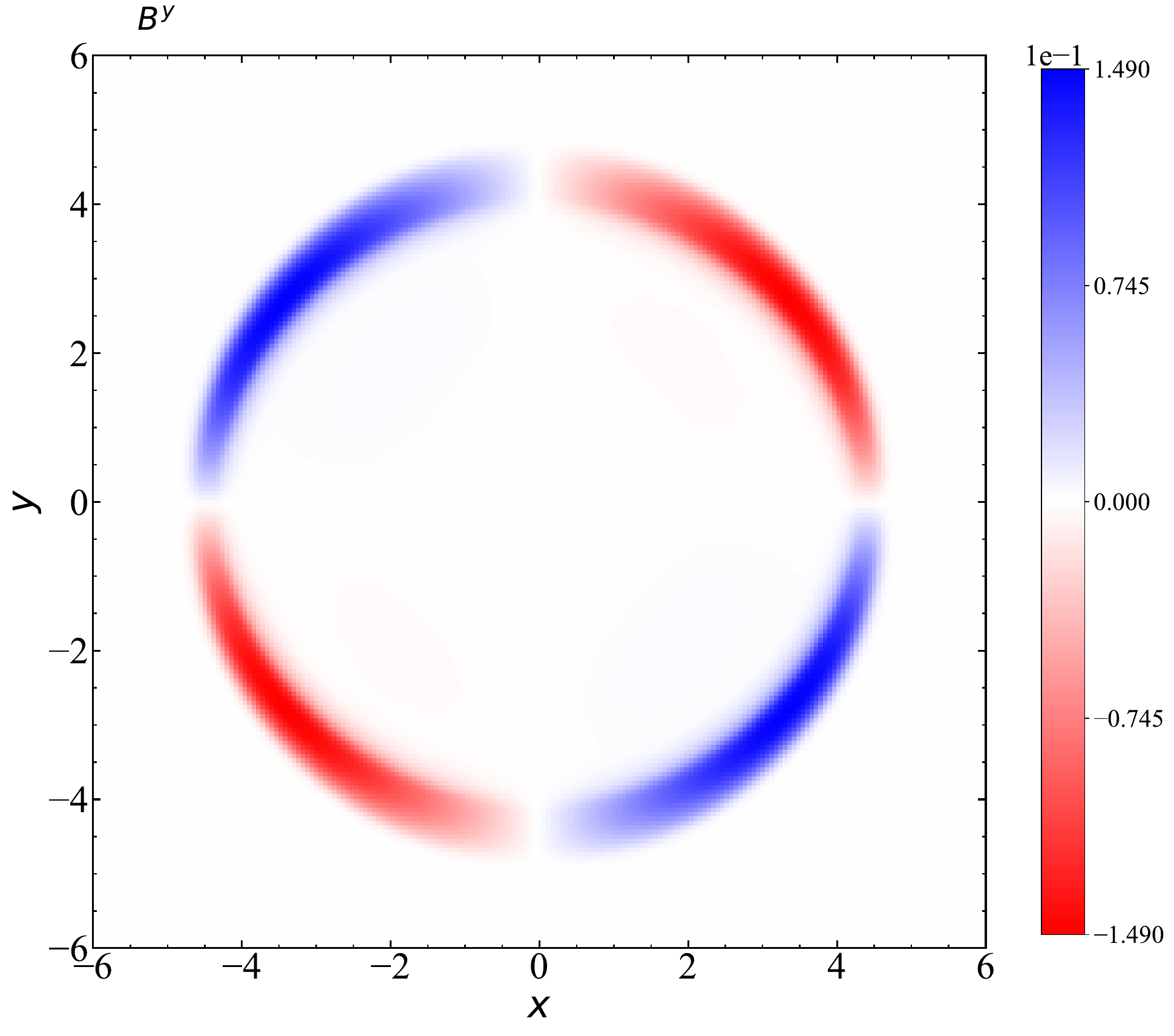}
     \caption{\small 
        The magnetic field components $B^x$ in the left panel and $B^y$ in the right panel at time $t = 4$ are depicted in this figure. A constant conductivity of \( \sigma = 10^6 \) is used.
    }
    \label{fig:cylexp}
\end{figure}

We use a grid of $200 \times 200$ cells in the x-y plane in the domain $[-6, 6]$, resulting in a resolution of $\Delta x = \Delta y = 0.06$. For this test, we use Neumann boundary conditions and TVD reconstruction with a minmod limiter, with the CFL set to $0.25$. Throughout the evolution, we keep the conductivity constant at $10^6$, and thus this test is expected to yield results close to those in the ideal MHD limit. We also observe similar results with lower conductivity values, such as $\sigma = 10^4$, with only negligible differences.
We use the following initial density profile to conduct the test in the x–y plane,

\begin{equation}
\rho(r) = 
\begin{cases}
\rho_{\text{in}}, & r \leq r_{\text{in}} , \\[1.2ex]
\exp\left[ 
\frac{(r_{\text{out}} - r)\ln \rho_{\text{out}} + (r - r_{\text{in}})\ln \rho_{\text{in}}}
     {r_{\text{out}} - r_{\text{in}}}
\right], & r_{\text{in}} < r < r_{\text{out}} , \\[2ex]
\rho_{\text{out}}, & r \geq r_{\text{out}} ,
\end{cases}
\end{equation}

\noindent where $r = \sqrt{x^2 + y^2}$ is the radial distance from the origin, and $r_{\text{in}}$ and $r_{\text{out}}$ are radial parameters. Inside a radius $r < 0.8$, we set the pressure to $p_{\text{in}} = 1$ and the density to $\rho_{\text{in}} = 0.01$. In the intermediate region $0.8 \leq r \leq 1.0$, both pressure and density decrease exponentially in the radially outward direction. In the outer region $r > 1$, we set the pressure and density as $p_{\text{out}} = \rho_{\text{out}} = 0.001$.
The magnetic field is uniform, and initially, the only non-trivial component is $B^x = 0.05$. In accordance with the ideal MHD approximation, we set the other quantities to zero. 
Fig.~\ref{fig:cylexp} illustrates the magnetic field components $B^x$ (left panel) and $B^y$ (right panel) at time $t = 4$.  Our findings are in good agreement with those reported in~\cite{komissarov1999, NEILSEN2008, DelZanna2007, Komissarov2007, Palenzuela2009, Cheong2022, Shankar2023}.
\subsection{Self-similar Current Sheet}
The self-similar current sheet is a standard test for evaluating magnetic diffusion in a plasma with finite electrical conductivity and was initially proposed by~\cite{Komissarov2007}. For this test, we fix the conductivity at $\sigma = 100$, representing a regime of moderate resistivity. We initialize only the magnetic field $B^y(x, t)$ and set $B^x(x, t) = B^z(x, t) = 0$, introducing a sign reversal across a thin current layer of width $\Delta L$. The system is initially in static equilibrium because the velocity field ($v^i = 0$) is set to zero, and the pressure and density are constant. 
The magnetic pressure is much smaller than the fluid pressure. Under these conditions, the magnetic field evolution follows a simple diffusion process described by the one-dimensional equation~\cite{Komissarov2007}:

\begin{align}
\partial_t B^y - \frac{1}{\sigma} \, \partial_x^2 B^y = 0.
\end{align}

We assume that the electric field and its temporal derivative are initially zero across the entire domain, i.e., $E^i = 0$ and $\partial_t E^i = 0$. The width of the current reversal layer increases significantly beyond $\Delta L$ as the system evolves, and it changes in a self-similar way. 

\begin{figure}[H]
    \centering
    \includegraphics[height=5.9cm, width=7.9cm]{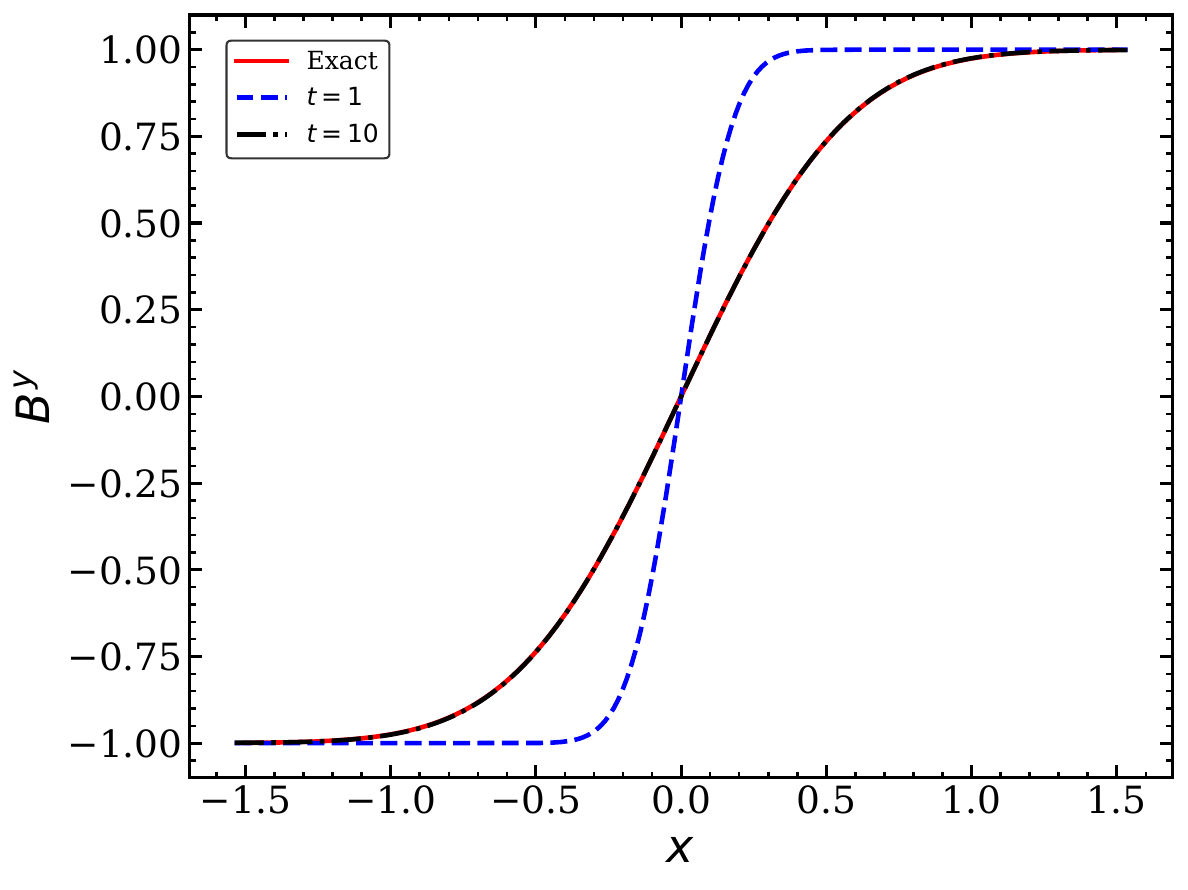}
     \caption{\small 
        This figure shows the magnetic field component $B^y$ as a function of $x$. Here, the conductivity is uniform, with a value of $\sigma = 10^2$, and the solution is calculated and displayed at $t = 1$ (initial time) and $t = 10$ with resolution $\Delta x = 0.005$. The initial time is set to $t = 1$ (rather than $t=0$) to avoid a singularity in the analytic solution.
    }
    \label{fig:current_sheet}
\end{figure}

The analytical exact solution is as follows for $t > 0$:
\begin{align}
B^y(x, t) = B_0 \, \mathrm{erf} \left( \frac{1}{2} \sqrt{\frac{\sigma}{\xi}} \right), \label{eq:current_sheet}
\end{align}
where "erf" is the error function and $\xi = t/x^2$.
We use a computational domain with extent \( x \in [-1.5, 1.5] \) and a resolution of \( \Delta x = 1/200 \), with the CFL number set to $0.25$. The initial pressure and density are set to \( p = 50 \) and \( \rho = 1 \), respectively. Here, \( t = 1 \) corresponds to the initial time, chosen instead of \( t = 0 \) to avoid the singular behavior in Eq.~\eqref{eq:current_sheet}. We perform the simulation until \( t = 10 \). As in the previous tests, we apply Neumann boundary conditions and TVD reconstruction with a minmod limiter. We show the exact solution, initial data, and final result of the numerical solution of the resistive MHD equations in Fig.~\ref{fig:current_sheet}. The exact and numerical solutions coincide almost perfectly, confirming that the magnetic field evolution at intermediate conductivity is accurately described by the resistive MHD equations. The results are consistent with~\cite{Komissarov2007, Palenzuela2009, Dionysopoulou2013, Cheong2022}.

\subsection{Large Amplitude Circularly Polarized Alfv\'{e}n Waves}
The large-amplitude circularly polarized (CP) Alfv\'{e}n wave test, introduced in one of the earlier relativistic MHD works by~\cite{DelZanna2007} and later employed in resistive and general-relativistic contexts~\cite{Palenzuela2009, Dionysopoulou2013}, examines the propagation of a nonlinear Alfv\'{e}n wave over a uniform background magnetic field $B_0$ within a domain with periodic boundary conditions. We use a computational domain of \( x \in [-0.5, 0.5] \), centered at $x = 0$, which corresponds to a domain length of $L_x = 1$ in the x-direction. Under these conditions, the wave returns to its initial location after one full period, $t = L_x/v_A = 2$. To recover the ideal MHD limit, we use a very high conductivity $\sigma = 10^6$. Similar to~\cite{DelZanna2007, Palenzuela2009, Dionysopoulou2013}, we determine the initial magnetic field and velocity components from the analytic solution of a circularly polarized Alfv\'{e}n wave, evaluated at $t=0$:

\begin{align}
&B^i = \left\{ B_0, \ \eta_A B_0 \cos[k(x - v_A t)], \ \eta_A B_0 \sin[k(x - v_A t)] \right\}, \label{eq:Bi} \\[1.2ex]
&v^i = -\frac{v_A}{B_0} (0, B^y, B^z), \label{eq:vyvz}
\end{align}

\noindent where $k = 2\pi$ is the wave vector, $\eta_A$ is the amplitude of the wave, and the Alfv\'{e}n velocity $v_A$ is defined by

\begin{equation}
v_A^2 = \frac{2 B_0^2}{h + B_0^2 (1 + \eta_A^2)}
\left[ 1 + \sqrt{1 - 
\left( \frac{2 \eta_A B_0^2}{h + B_0^2 (1 + \eta_A^2)} \right)^2 } \right]^{-1}. \label{eq:Alf_vel}
\end{equation}

\noindent Given the parameters $\rho = p = \eta_A = 1$ and $B_0 = 1.1547$, we calculate the Alfv\'{e}n velocity to be $v_A = 0.5$. The following equation provides the initial electric field~\cite{DelZanna2007}:

\begin{align}
\vec{E} = -\vec{v} \times \vec{B} = (0,v_A B_z, - v_A B_y).
\end{align}

We apply periodic boundary conditions and compare the evolved propagating wave profile after one full period with the initial profile to ensure the accuracy of the code. In order to evolve the system, we use TVD reconstruction with a monotonized central-difference limiter of the second order (MC2), and set the CFL number to $0.25$. The resistive MHD equations enable us to recover the ideal MHD limit by considering very high conductivity and increasing resolution. 

\begin{figure}[H]
    \centering
    \includegraphics[height=5.9cm, width=7.9cm]{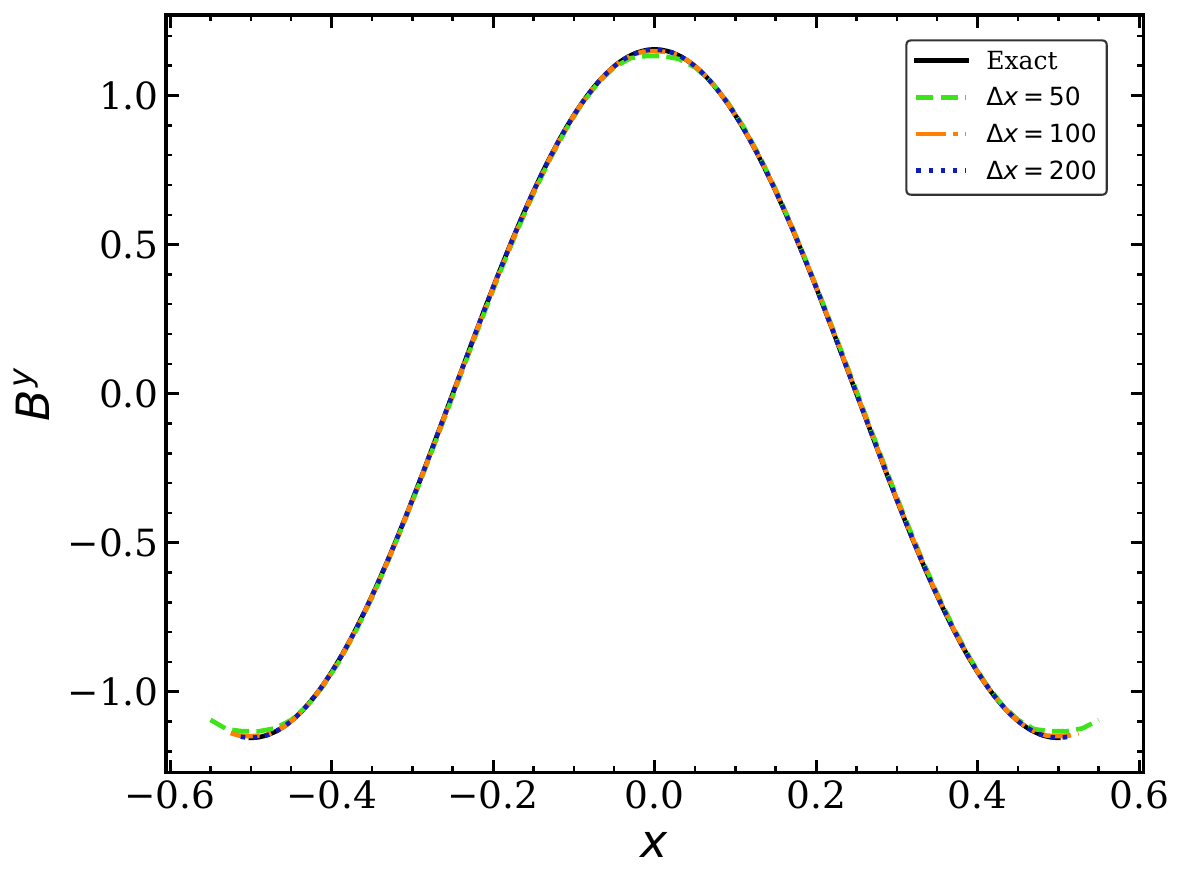}
     \caption{\small 
        The $B^y$ component of the magnetic field for three different spatial resolutions $\Delta x = \{1/50, 1/100, 1/200\}$, along with the exact solution depicted by the solid black line.
    }
    \label{fig:Alfven}
\end{figure}

We conduct three simulations with a uniform conductivity of $\sigma = 10^6$ using the following resolutions \( \Delta x = \left\{ 1/50, 1/100, 1/200 \right\} \). To measure the error, we compare the numerical solution after one period at \( t = 2 \) with the initial condition at \( t = 0 \). Consistent with the initial conditions at \( t = 0 \), Fig.~\ref{fig:Alfven} shows the \( B^y \) component at three different resolutions after a full cycle at \( t = 2 \). 
As mentioned earlier, the higher the resolution, the closer the final solution with high conductivity at \( t = 2 \) is to the exact solution, indicating that the numerical solution of the resistive MHD equations under these conditions tends toward the ideal MHD limit. Our results are in full agreement with~\cite{Palenzuela2009, Dionysopoulou2013}.

\subsection{3D TOV Star}
In this section, we simulate the evolution of a stationary, spherically symmetric neutron star in the presence of electromagnetic fields in three dimensions. In this test, no magnetosphere is formed because the star is non-rotating and no electric charges are present. Consequently, the exterior of the star can be considered an electrovacuum. Since this test is performed in three dimensions and involves both general relativity and resistive magnetohydrodynamics, it is more challenging than previous tests. This is our first test in a dynamical spacetime, where we evolve the spacetime variables using the Z4c formalism.

\subsubsection{3D TOV star with confined magnetic fields}

We obtain the neutron star model by solving the TOV equation~\cite{Tolman1939}, and then add a poloidal magnetic field confined to the stellar interior~\cite{Duez_1_2006, Duez_2_2006}. To obtain a poloidal magnetic field, we define a vector potential as follows~\cite{Palenzuela2009}:

\begin{align}
A_{\phi} = \varpi^{2} \max\left[ A_{b} \left( P - P_{\mathrm{cut}} \right), 0 \right]^{2},
\end{align}

where $\varpi \equiv \sqrt{(x - x_{\star})^{2} + y^{2}}$ is the cylindrical radius, $A_b$ specifies the initial magnetic field strength, and $(x_{\star},0,0)$ is the location of the stellar center in the computational domain. $P_{\text{cut}}$ is the cut-off pressure, equal to $4\%$ of the central pressure, below which the magnetic field is set to zero. In this setup, $A_b \approx 1.58$ and $P_{\text{cut}} \approx 6.53 \times 10^{-6}$.
We set up the initial data for the fluid and the spacetime variables such that they satisfy the Einstein field equations. We then superimpose the magnetic field onto the fluid, and evolve the magnetohydrodynamics variables using the resistive version of \texttt{GRaM-X}, while we evolve the spacetime variables using the Z4c formulation. We set the constraint damping parameters to \( \kappa_1 = 0.02 \) and \( \kappa_2 = 0.0 \), and we fix the dissipation coefficient at \( 0.32 \)~\cite{Shankar2023}.

We produce a TOV star with a mass of $1.4\,M_\odot$ and a radius of $8.125\,M_\odot$ by solving the one-dimensional TOV equations using a polytropic EoS with parameters $K = 100$, $\Gamma = 2$, and an initial central density of $1.28 \times 10^{-3} \, M_\odot^{-2}$. We then interpolate this 1D initial data onto a 3D computational grid, and use it as the initial setup for our simulation.
Combined with the above magnetic field prescription, this setup yields a poloidal magnetic field with a strength of $B_c \approx 4.3 \times 10^{-8}$ at the center of the star and a maximum pressure of $P_{\text{max}} \approx 1.6 \times 10^{-4}$ in code units. From these values, the ratio of fluid pressure to magnetic pressure is given by $\beta = \frac{2P_{max}}{B_{max}^2} \approx 1.8 \times 10^{11}$. This ratio indicates that the magnetic field has a negligible impact on the static configuration of the star over the relevant timescales. For all subsequent evolutions, we use an ideal-gas EoS with $\Gamma = 2$. 

To recover the ideal MHD limit in the deep interior of the star, where the conductivity is extremely high, and the electrovacuum limit in the exterior, where both density and isotropic conductivity are negligible, we define a spatially-varying conductivity. This approach automatically enforces the correct boundary conditions at the stellar surface, similar to previous methods~\cite{Baumgarte2003, Lehner2012}, but without the need for an analytical interior solution or the complications of matching the interior and exterior electromagnetic fields. In our simulations, we implement this by defining the conductivity as a function of the conserved rest-mass density. This approach ensures a smooth and continuous transition from the highly conductive interior to the electrovacuum exterior~\cite{Palenzuela2009}:

\begin{align}
\sigma = \sigma_{0} \max\left[ \left( 1 - \frac{D_{\mathrm{atmo}}}{D} \right), 0 \right]^{2}.
\label{eq:dynamo_sigma}
\end{align}

In regions with high rest-mass density, such as in the stellar core, we set $\sigma = \sigma_0$. To model the exterior of the star, we employ an artificial atmosphere where the density is set to $10^{-10}\,M_\odot$ and the fluid velocity is zero. When $D = D_{\text{atmo}}$, the conductivity in the atmosphere becomes zero. We also set the initial electric and magnetic fields in the atmosphere to zero. With zero conductivity, the evolution of the electric and magnetic fields in the atmosphere is then governed by Maxwell’s equations in the absence of electric currents. In this region, we determine the pressure and specific energy density using a polytropic EoS with $K = 100$ and $\Gamma = 2$. 
We perform three simulations, each using four AMR levels. 
The finest spatial resolutions are \( \Delta x = \{0.3\,M_\odot,\ 0.2\,M_\odot,\ 0.15\,M_\odot\} \), 
corresponding, respectively, to the number of grid points across the finest level 
\( N = \{80,\ 120,\ 160\} \).
The outermost cube extends to $\pm 96$ for $\Delta x = 0.3\,M_\odot$ and $\Delta x = 0.2\,M_\odot$, and to $\pm 120$ for the $\Delta x = 0.15\,M_\odot$ case. We run the finest-resolution simulation using a larger domain (\( \pm 120 \)) to avoid boundary effects during the longer evolution. The extent of the innermost cube is $\pm 12$ in all three simulations.
We carry out all simulations in full 3D without any symmetries. In this test, we use the HLLE Riemann solver with fifth-order WENO reconstruction, and the CFL factor is set to $0.125$. We consider different values of $\sigma_0$ = \{$10^6, 10^3, 200, 100\}$.

\begin{figure}[H]
    \centering
    \includegraphics[height=7.64cm, width=8.64cm]{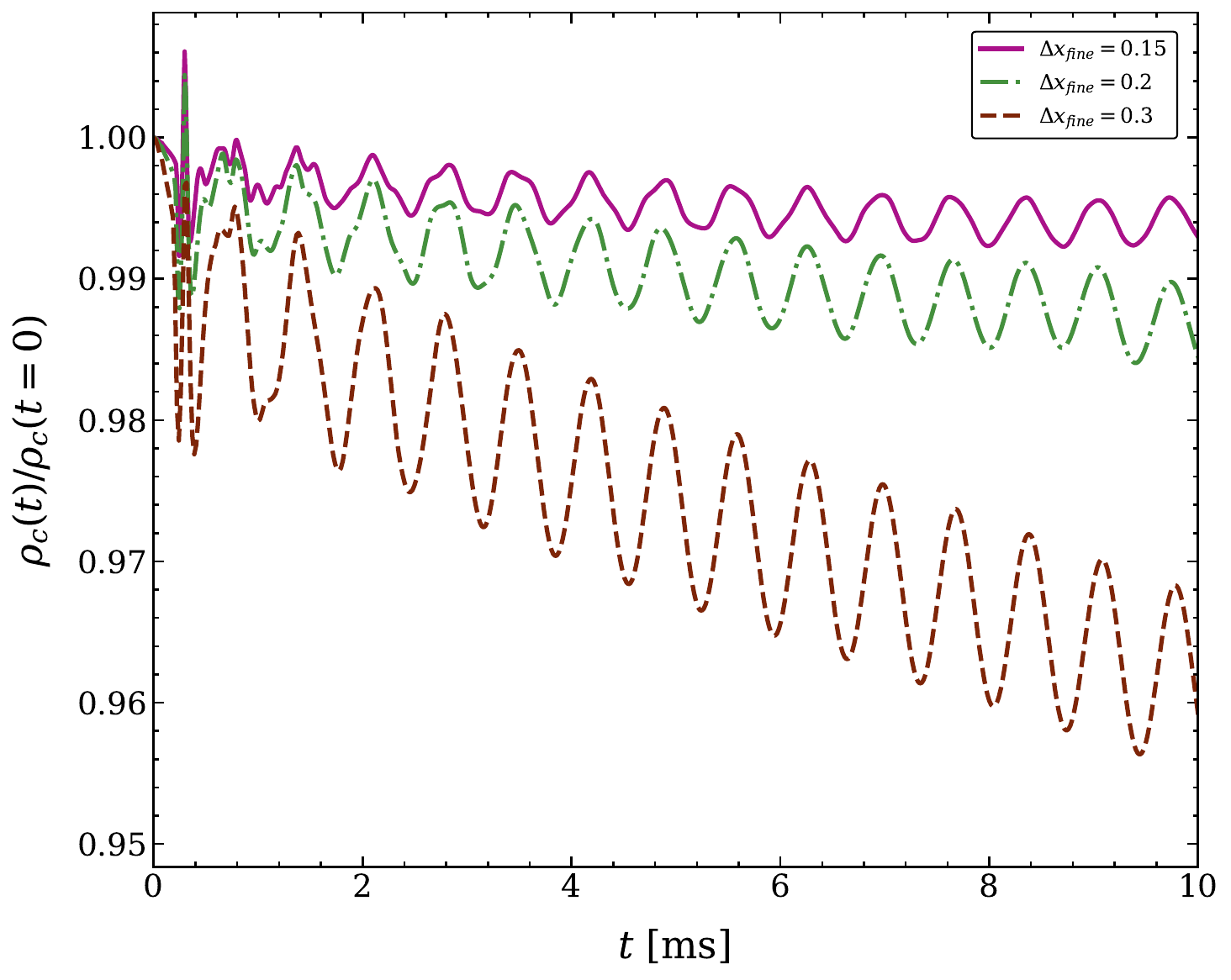}
    \hspace{0.3cm}
    \includegraphics[height=7.8cm]{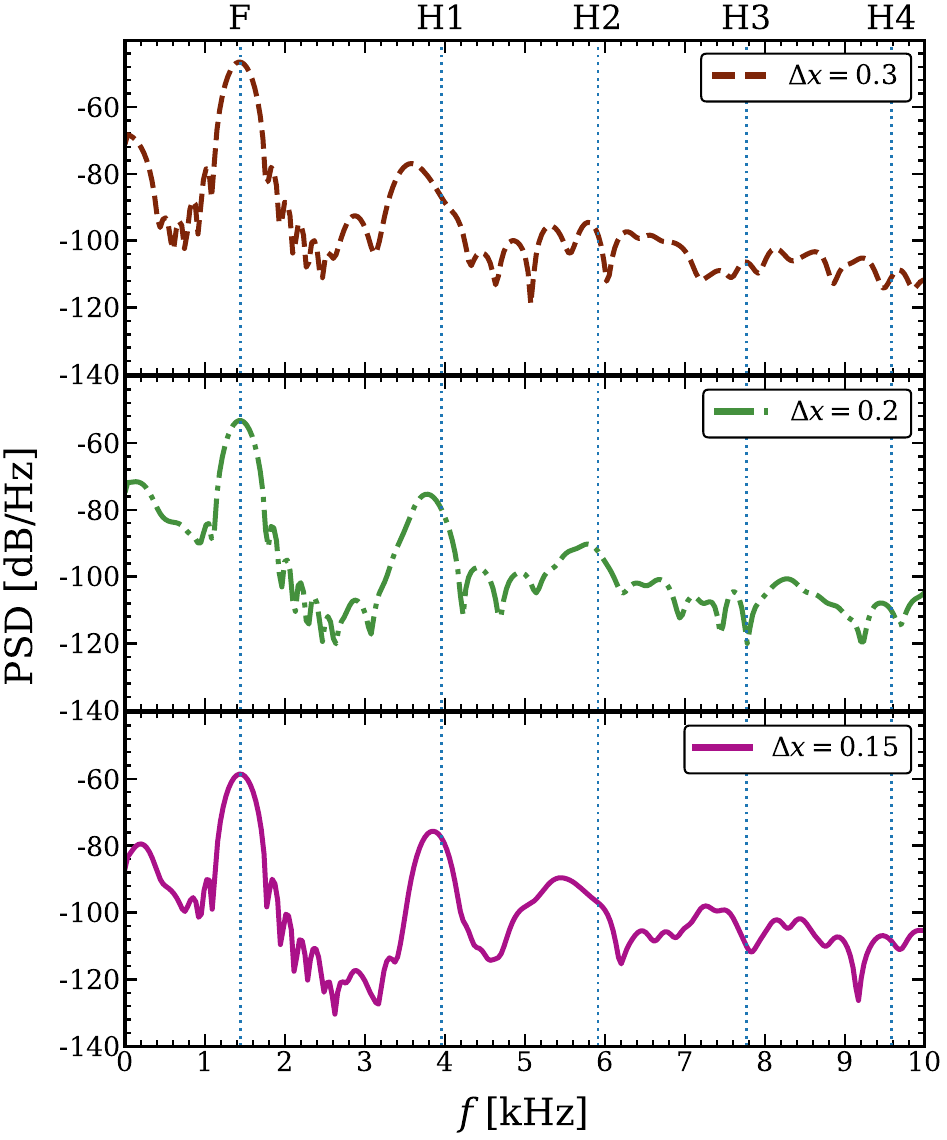}
     \caption{\small 
        Left panel: Normalized central density as a function of time for a non-rotating magnetic star in dynamical spacetime. Right panel: PSD of central density oscillations for a stable star with confined magnetic fields.
        Different line styles indicate different resolutions: brown dashed line $\Delta x = 0.3\,M_\odot$, green dash-dotted line $\Delta x = 0.2\,M_\odot$, and the dark pink solid line $\Delta x = 0.15\,M_\odot$. 
    }
    \label{fig:rho_PSD}
\end{figure}

Our initial data consists of a static TOV star which is expected to remain in a stationary configuration when evolved in time (the ratio of fluid pressure to magnetic pressure is $1.8\times10^{11}$, hence magnetic fields, even though present, are dynamically unimportant in the evolution). However, the finite numerical resolution cannot represent this equilibrium perfectly. The resulting truncation error acts as a small perturbation to the stellar configuration, which triggers oscillations around equilibrium. 
The left panel of Fig.~\ref{fig:rho_PSD} shows the time evolution of the central rest-mass density at three different resolutions with fixed \( \sigma_0 = 10^6 \), normalized by its initial value \( \rho_{c,0} \), in a dynamical spacetime. 

As the resolution increases, the truncation error is reduced, leading to a corresponding decrease in the amplitude of the oscillations, and the central density converges to the equilibrium configuration.
These frequencies are different between static and dynamical spacetime evolutions because, in the static case, the spacetime metric remains fixed while only the fluid variables evolve. This means that we can extract only the fluid oscillation modes. 
In contrast, when the spacetime evolves dynamically, both the matter and metric fields respond to perturbations, giving rise to different frequencies of oscillation. 
As a result, modes linked to the dynamics of gravity can be captured only when the evolution is fully dynamical~\cite{Shankar2023}. 

The frequencies of the central density oscillations can be measured by performing a Fourier transform of their time variations. The right panel of Fig.~\ref{fig:rho_PSD} displays the power spectral density (PSD) of the central density oscillations at three different resolutions. The dotted vertical lines represent the eigenfrequencies predicted by linear perturbation theory \cite{Yoshida2001, Font2002}. At the lowest resolution, $\Delta x = 0.3\,M_\odot$, we only resolve the fundamental mode $F$.
At higher resolutions, we also resolve the first harmonic $H1$. It is expected that, with further increases in resolution and consequently in computational cost, higher modes will also be resolved.

\begin{figure}[H]
    \centering
    \includegraphics[height=5.9cm, width=7.9cm]{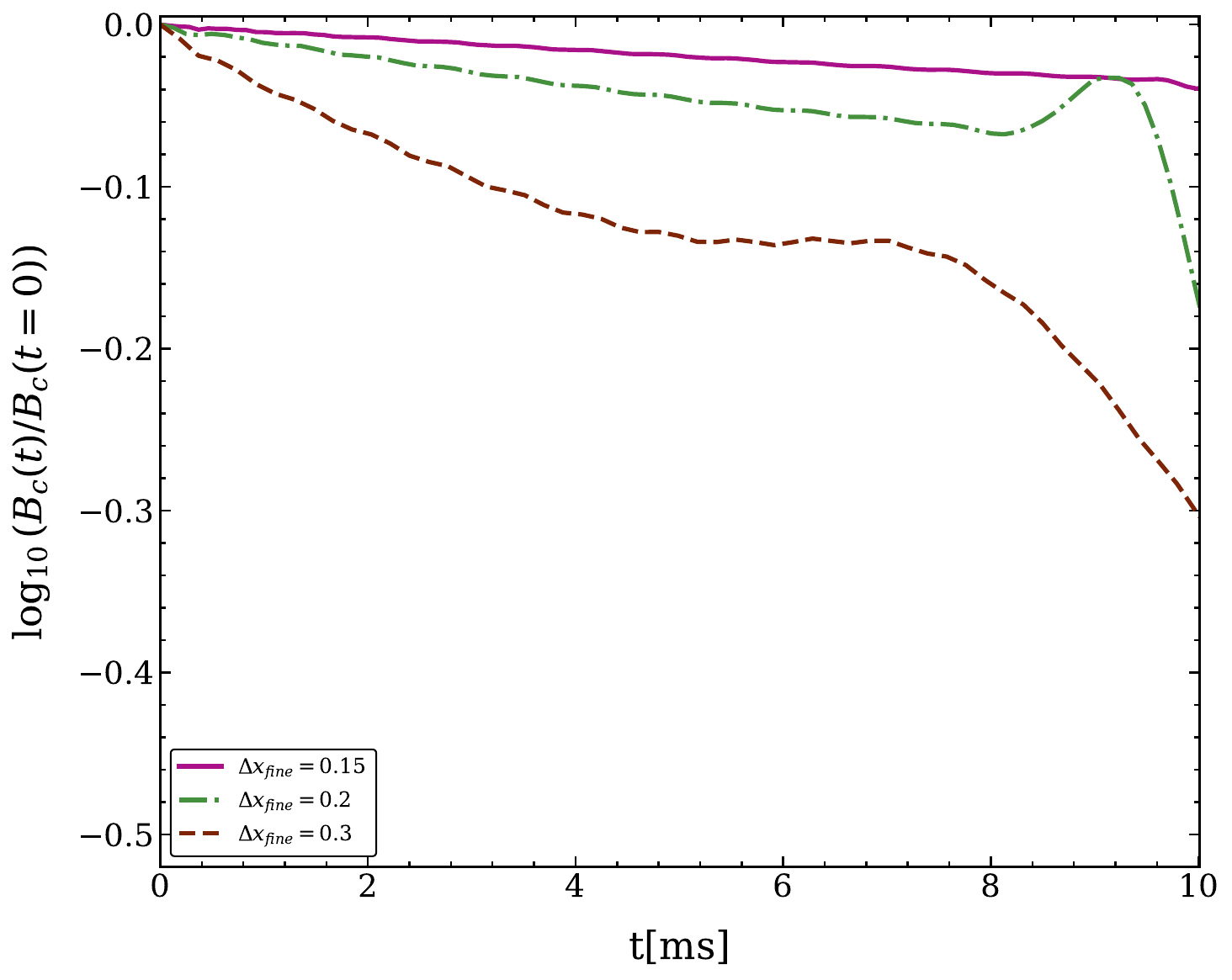}
    \hspace{0.1cm}
    \includegraphics[height=5.9cm, width=7.9cm]{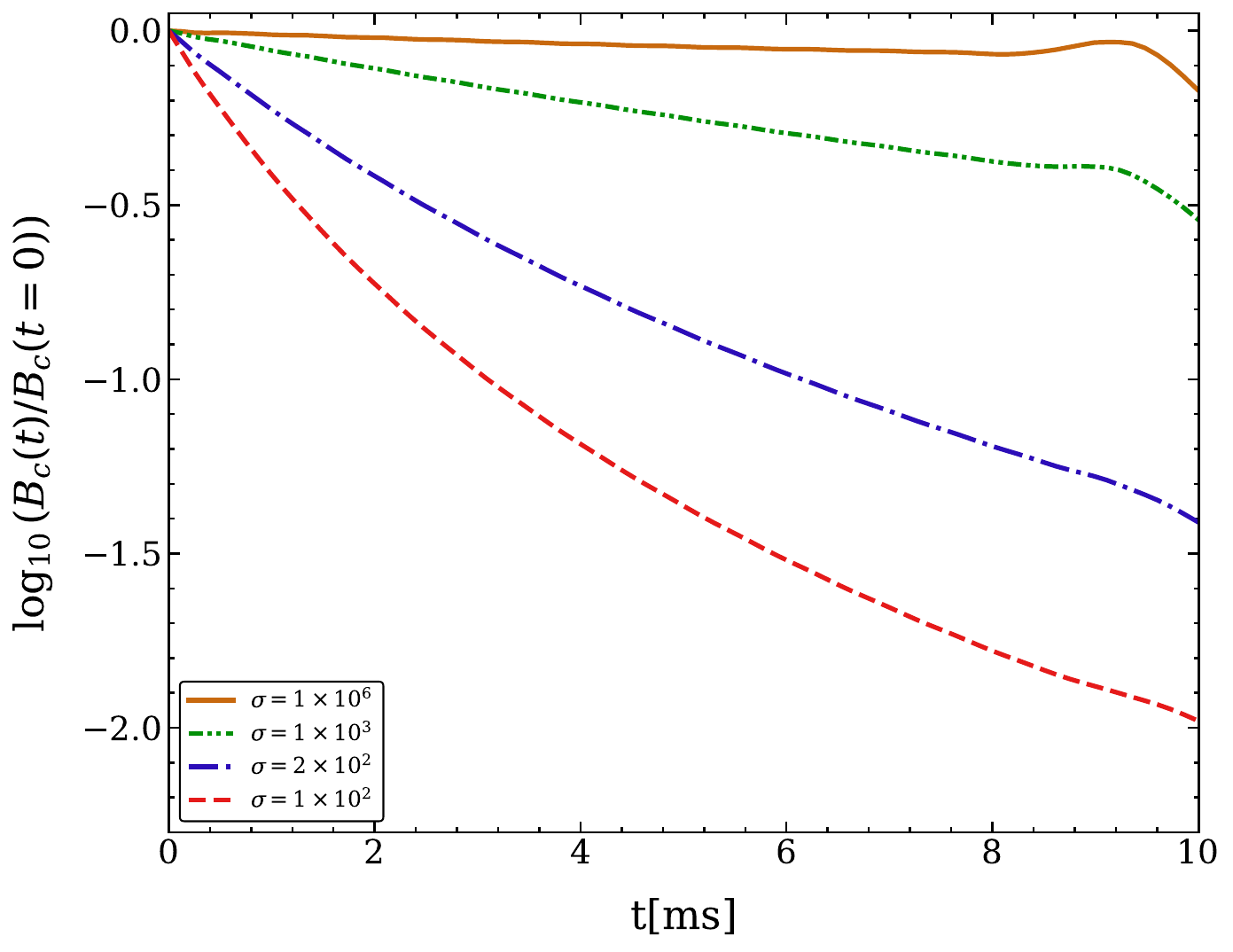}
     \caption{\small 
        Left panel: Normalized central magnetic field as a function of time for a non-rotating magnetic star in dynamical spacetime for different resolutions with \( \sigma_0 = 10^6 \): brown dashed line $\Delta x = 0.3\,M_\odot$, green dash-dotted line $\Delta x = 0.2\,M_\odot$, and dark pink solid line $\Delta x = 0.15\,M_\odot$. Right panel: Time evolution of the central magnetic field for conductivities ranging from $\sigma = 10^6$ to $\sigma = 10^2$, computed at an intermediate resolution of $\Delta x = 0.2\,M_\odot$.
    }
    \label{fig:Bc_plots}
\end{figure}

The left panel of Fig.~\ref{fig:Bc_plots} shows the time evolution of the central magnetic field at three different resolutions with fixed \( \sigma_0 = 10^6 \). Similar to the central rest-mass density, the central magnetic field gradually decreases during the evolution. This slow drift toward lower values arises mainly from intrinsic numerical resistivity, which decreases with increasing resolution.
To distinguish numerical dissipation from physical resistive effects, we carry out additional simulations including explicit finite conductivities, with values ranging from $\sigma_0 = \{ 10^6, 10^3, 2 \times 10^2, 10^2\}$, using a medium resolution of $\Delta x = 0.2\,M_\odot$. By introducing finite conductivity, we can directly model the physical Ohmic diffusion of the magnetic field, rather than the artificial decay that arises from limited numerical resolution.
At this and all other resolutions, the fluid velocity remains negligible and can be effectively considered zero.
As a result, the evolution of the central magnetic field can be described by a simple diffusion process. The associated Ohmic decay time is inversely proportional to the conductivity. 

As shown in the right panel of Fig.~\ref{fig:Bc_plots}, the results for different conductivities converge to an exponential decay, consistent with the expected resistive behavior. In particular, for $\sigma_0 = 10^2$, the magnetic field decreases by almost two orders of magnitude within 10 ms.
We note that, although the velocity in the atmosphere is set to zero, over long evolution times this artificial constraint produces side effects, such as an increase in the central magnetic field. Physically, such a growth should not occur. During the system’s evolution, the outer layers of the star slowly expand while the inner layers shift inward toward the center. In the ideal MHD approximation, magnetic field lines are “frozen” into the fluid and move along with it. As the fluid approaches the stellar center, the field lines are compressed near the core. This displacement process artificially amplifies the strength of the central magnetic field, even though the total magnetic energy (integrated over the entire star) remains constant. Therefore, this is not a genuine amplification, but rather a numerical artifact caused by the treatment of the atmosphere. As a result, imposing zero velocity in the atmosphere is only valid and the results remain reliable as long as no artificial growth of the magnetic field takes place. 

\begin{figure}[H]
    \centering
    \includegraphics[height=6.5cm, width=8.5cm]{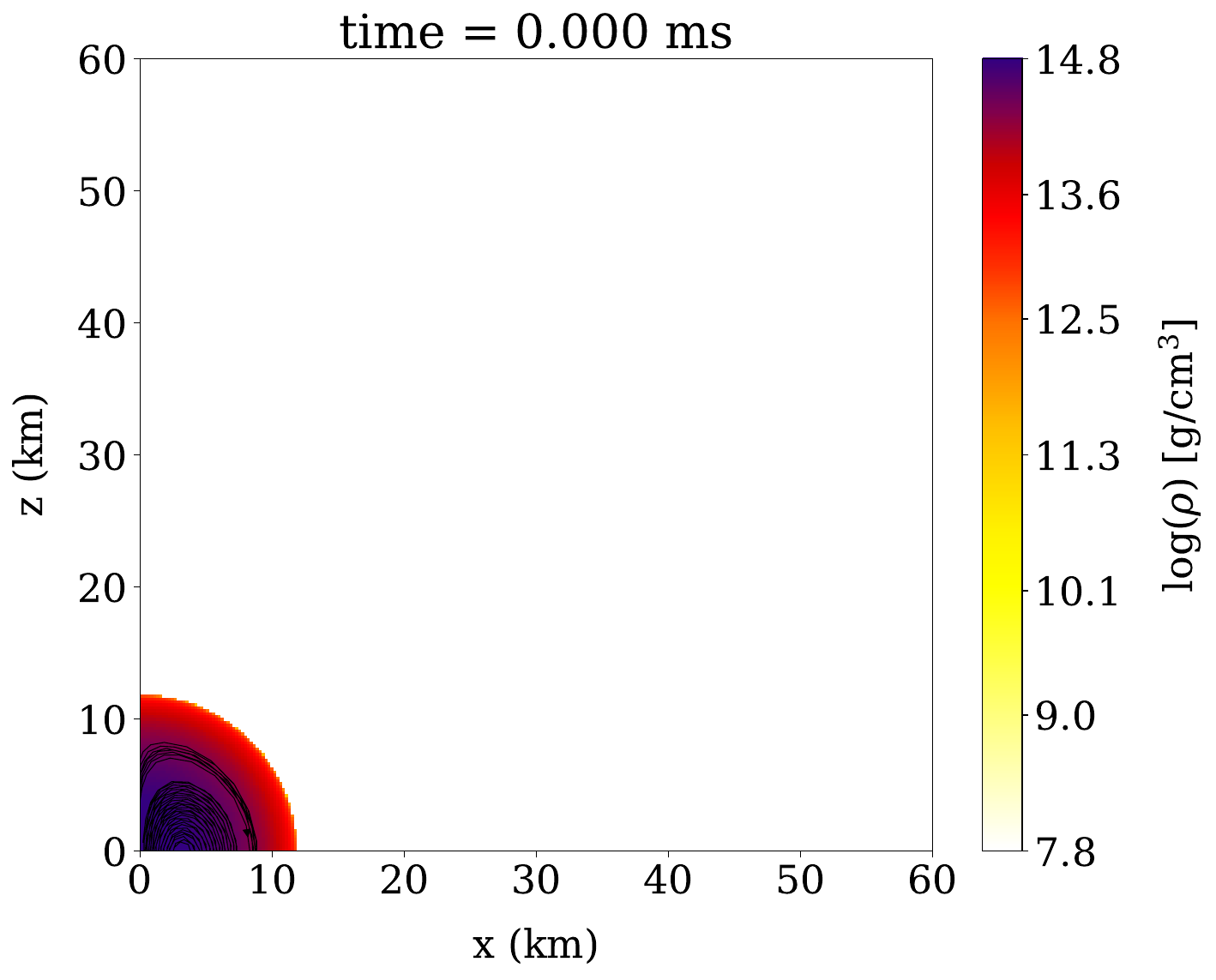}
    \hspace{0.1cm}
    \includegraphics[height=6.5cm, width=8.5cm]{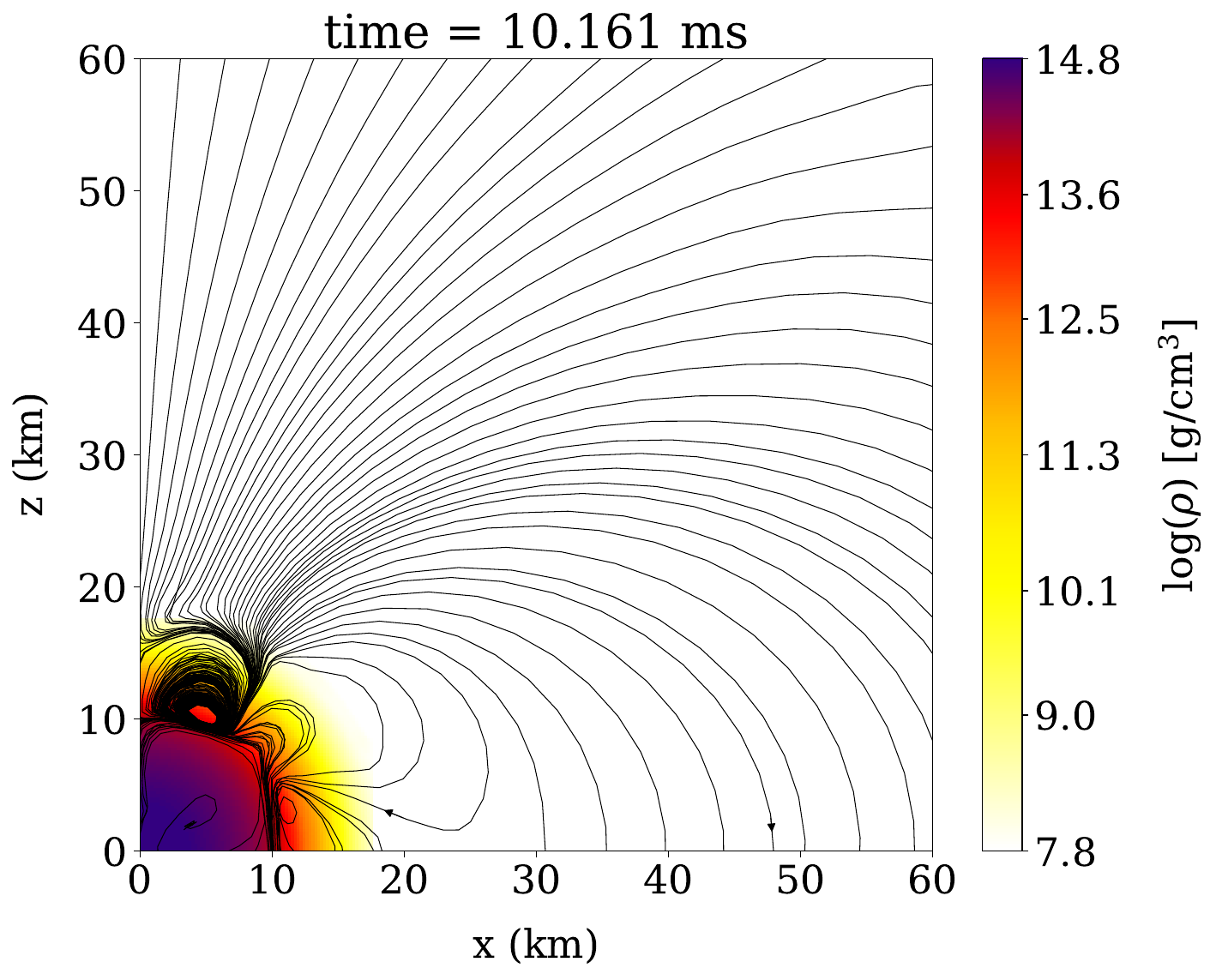}
     \caption{\small 
        Two-dimensional slices in the $(x, z)$ plane show the rest-mass density and magnetic field lines with \( \Delta x = 0.15 \) and fixed \( \sigma_0 = 10^6 \) at $t = 0 \,\text{ms}$ and $t = 10.161 \,\text{ms}$. The left panel illustrates a non-rotating star with its initial magnetic field confined entirely within the stellar interior at $t = 0 \,\text{ms}$. The right panel demonstrates that, as the star evolves over time, both numerical and physical resistivity cause the magnetic field to diffuse outward from the stellar interior.
    }
    \label{fig:BF-streamline}
\end{figure}

The left panel of Fig.~\ref{fig:Bc_plots} shows that, in the worst case, seen in the lowest-resolution test, the results remain reliable up to about 7 ms, after which artificial growth of the magnetic field sets in. For medium and high resolutions, this artificial growth occurs later, at approximately 8 and 9 ms, respectively.
Fig.~\ref{fig:BF-streamline} shows the rest-mass density and magnetic field lines of a non-rotating oscillating star in a two-dimensional (\( x, z \)) slice, with a resolution of \( \Delta x = 0.15 \) and fixed \( \sigma_0 = 10^6 \).
The left panel shows the initial rest-mass density and the magnetic field, which is confined to the stellar interior at time zero. 

As the time evolution begins, the magnetic field slowly leaks from inside the star to the exterior, eventually filling the computational domain. This process occurs because the system possesses a small but finite numerical resistivity, together with a non-zero physical conductivity near the stellar surface. As we have mentioned, the resistivity outside the star is set to zero, and Maxwell's equations reduce to the vacuum equations. In the stellar exterior, where the conductivity is set to zero and Maxwell’s equations reduce to their vacuum form, the external magnetic field is divergence-free. Under these conditions, the Laplacian of the field is negligible, allowing the field to be treated with high accuracy as a potential field, which is consistent with the observed dipole-like structure. As we increase the resolution of the simulation, the magnetic field relaxes to a stationary dipolar-like structure over longer timescales~\cite{Dionysopoulou2013}. 

In \cite{Dionysopoulou2013}, to avoid complications arising from large current gradients, they compute the charge density as \( q = \nabla_i E^i \) instead of evolving it directly. 
To assess the impact of this approach on our results, we performed an additional test using the same method. The results obtained show no significant differences compared to our standard evolution.
The structure of the magnetic field lines shown in the right panel of Fig.~\ref{fig:BF-streamline} near the stellar surface differs from the results reported in~\cite{Dionysopoulou2013}. However, farther away from the surface, the magnetic field lines remain stable, and their configuration agrees with~\cite{Dionysopoulou2013}.
This discrepancy arises from differences in the choice of numerical schemes, such as atmosphere treatment, the resolution, and the definition of characteristic speeds in the Riemann solver, which can lead to enhanced diffusion of the magnetic field lines close to the stellar surface. 

To further investigate how the choice of numerical scheme affects the results, we modified the definition of the characteristic speed in the Riemann solver. Specifically, we set the characteristic speed to match the value used in ideal GRMHD inside the star, while outside it is fixed to the speed of light. 
The results obtained with this configuration show slight differences from those in Fig.~\ref{fig:BF-streamline}, with reduced magnetic-field diffusion near the stellar surface. 
This demonstrates the importance of defining a characteristic speed that is consistent with the physical system to achieve higher accuracy in the simulations.

\subsubsection{3D TOV star with extended magnetic fields}

In this section, we examine a spherical magnetized star with a poloidal magnetic field extending beyond the stellar surface. We compute the initial magnetic field using the following vector potential:

\begin{equation}
A_r = A_\theta = 0, \qquad 
A_\phi = \frac{A_b}{2} \, \frac{r_0^{3}}{r^{3} + r_0^{3}} \, r \sin\theta .
\label{eq:poloidal_BF} 
\end{equation}

\noindent where $A_{r,\theta,\phi}$ denotes the components of the vector potential in the $r$, $\theta$, and $\phi$ directions, $r$ is the radial coordinate, and $A_b$ sets the initial magnetic field strength. In this setup, we choose \( r_0 = 5\,M_\odot \). By adopting \( \Gamma = 2 \) and \( K = 372 \), we model the non-rotating star as a polytrope. With these parameters, the star has a gravitational mass of $M = 1.33\,M_\odot$ and a radius of $20.68\,M_\odot$.  In this test, we use the same atmosphere treatment as in the previous section to recover the ideal MHD limit in the deep interior of the star and the electrovacuum limit in the exterior. 
The conductivity is computed from Eq.~\eqref{eq:dynamo_sigma} with a fixed value of \( \sigma_0 = 10^6 \).

\begin{figure}[H]
    \centering
    \includegraphics[height=6.5cm, width=8.5cm]{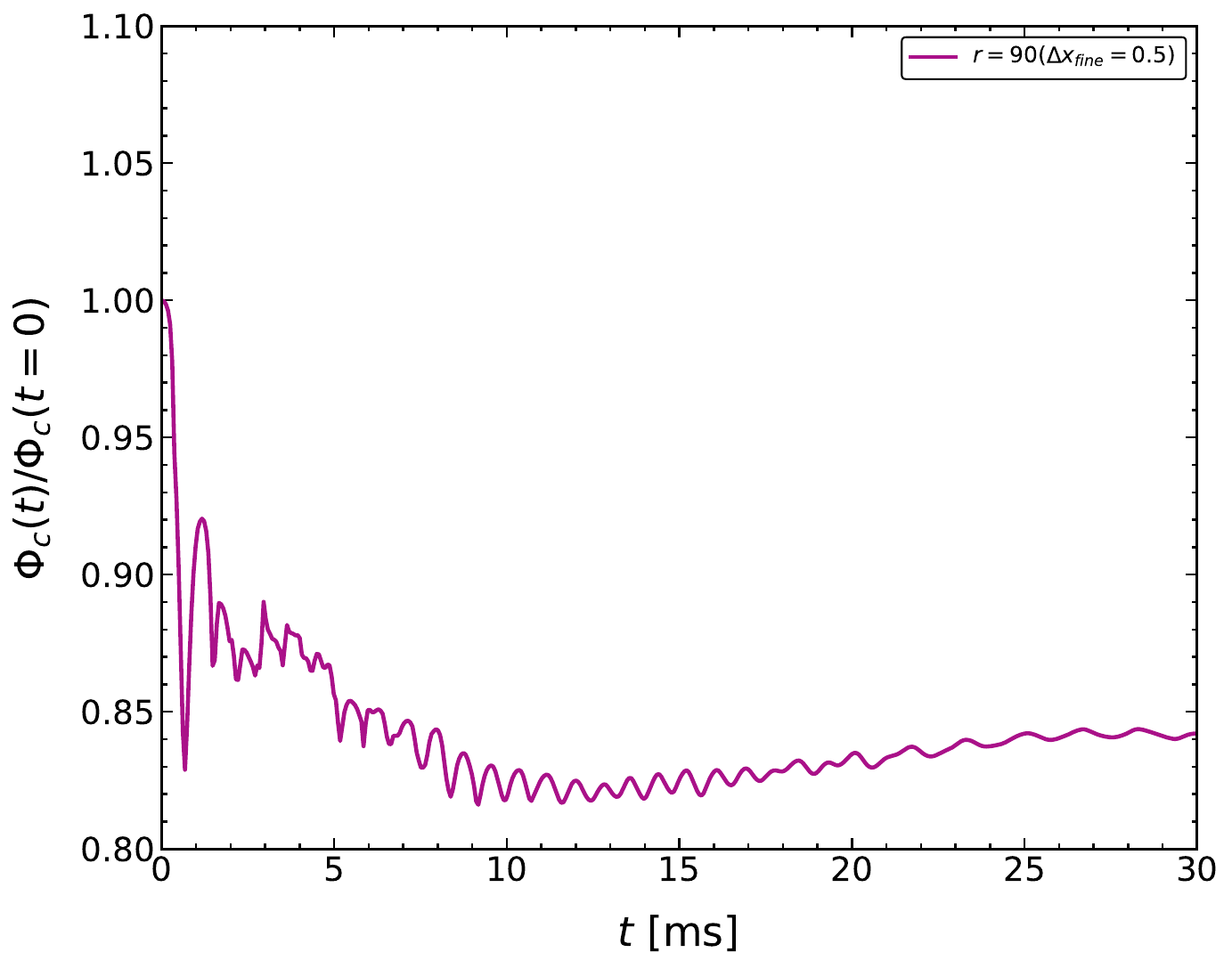}
    \hspace{0.1cm}
    \includegraphics[height=6.5cm, width=8.5cm]{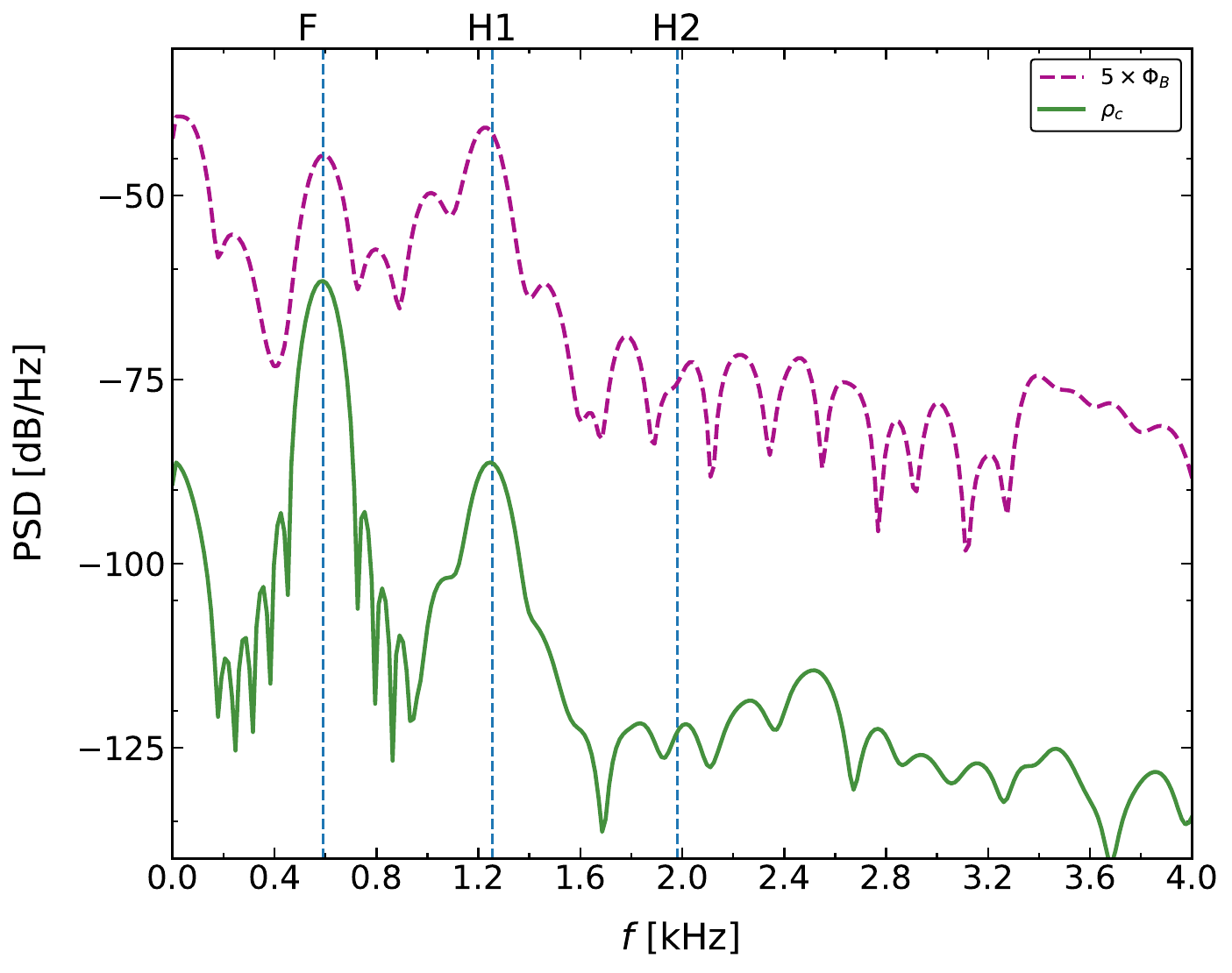}
     \caption{\small 
        Left panel: Evolution of the normalized magnetic flux as a function of time in a dynamical spacetime. The magnetic flux is computed on a hemispheric surface with radius $r=90\,M_\odot$. Right panel: Power spectral density of the rest-mass density and magnetic flux from 8 ms to 38 ms. The system reaches a quasi-equilibrium state after 8 ms, and the results become more stable.
    }
    \label{fig:BF-Flux}
\end{figure}

The strength of the initial polar magnetic field is $B_c = 1.02\times10^{-5}$.
We use four AMR levels, each with 120 cells. The computational domain extends to an outermost cube of $\pm 240 \,M_\odot$ and an innermost cube of $\pm 30\,M_\odot$. The coarsest and finest resolutions are $\Delta x = 4\,M_\odot$ and $\Delta x = 0.5\,M_\odot$, respectively. We use Neumann boundary conditions and WENO reconstruction, with the CFL number set to 0.125.
In this test, as briefly described in the previous section, we used a different characteristic speed in the HLLE Riemann solver than in the previous experiments: it is set equal to the ideal GRMHD value inside the star and to the speed of light outside the star. This choice improves system stability and reduces diffusion near the stellar surface.

We show the evolution of the magnetic flux through a hemispheric surface of radius \( r = 90\,M_\odot \) in the left panel of Fig.~\ref{fig:BF-Flux}. 
The results exhibit oscillations originating from variations in the rest-mass density~\cite{Dionysopoulou2013}.
The oscillations in the magnetic flux indicate that the star is not initially in equilibrium.  After approximately \( 8\,\mathrm{ms} \), the star reaches a quasi-equilibrium state.  The right panel of Fig.~\ref{fig:BF-Flux} shows the PSD of the magnetic flux and central rest-mass density from \( 8\,\mathrm{ms} \) to the final time \( t = 38\,\mathrm{ms} \). The excellent agreement between the power spectrum of the magnetic flux evolution and the corresponding spectrum of the central rest-mass density clearly confirms this behavior. These findings are consistent with previous studies, which report the same oscillatory behavior in magnetic flux and rest mass density~\cite{Dionysopoulou2013}.

\begin{figure}[H]
    \centering
    \begin{subfigure}[b]{0.48\textwidth}
        \centering
        \includegraphics[width=\textwidth]{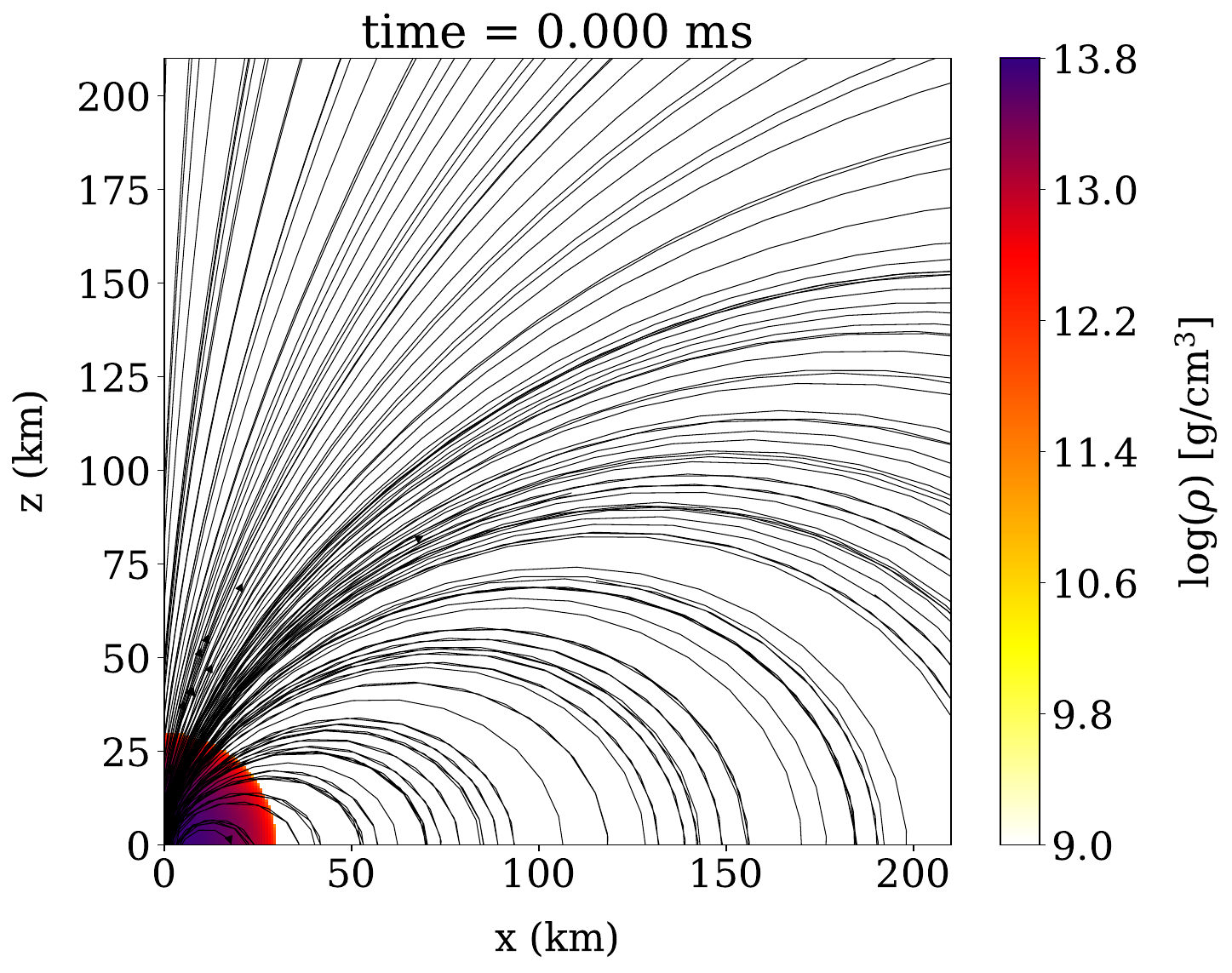}
    \end{subfigure}
    \hspace{0.1cm}
    \begin{subfigure}[b]{0.48\textwidth}
        \centering
        \includegraphics[width=\textwidth]{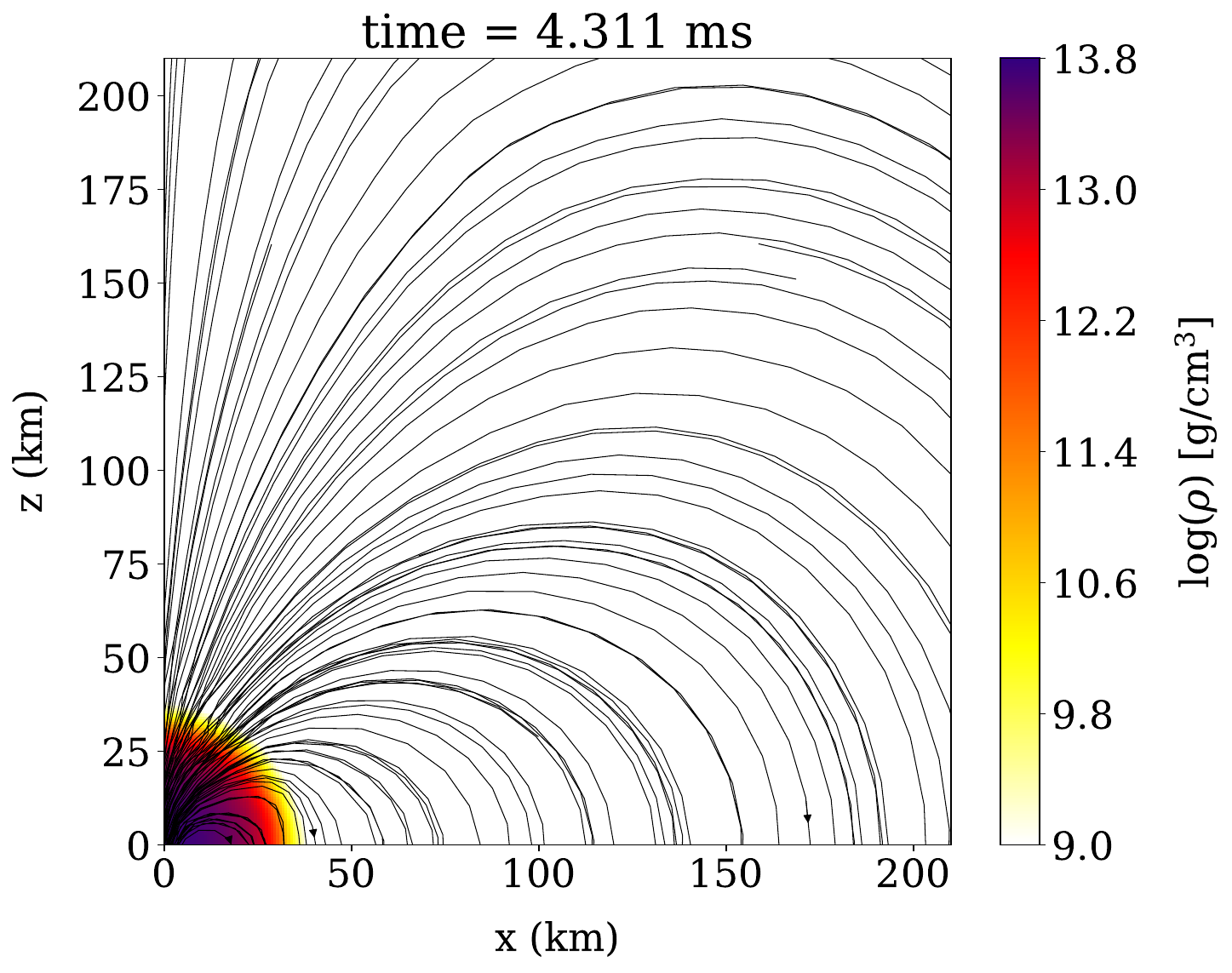}
    \end{subfigure}
    
    \begin{subfigure}[b]{0.48\textwidth}
        \centering
        \includegraphics[width=\textwidth]{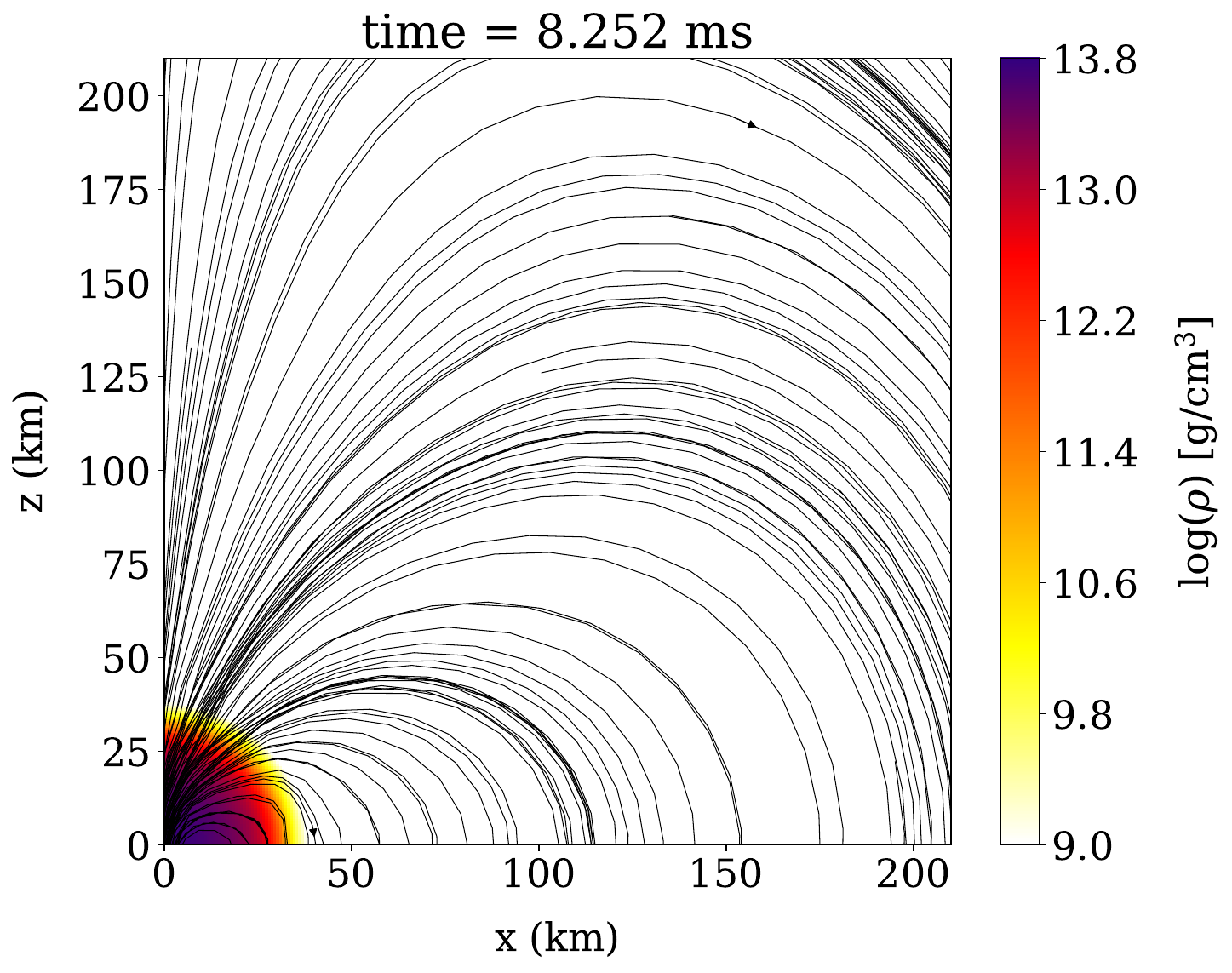}
    \end{subfigure}
    \hspace{0.1cm}
    \begin{subfigure}[b]{0.48\textwidth}
        \centering
        \includegraphics[width=\textwidth]{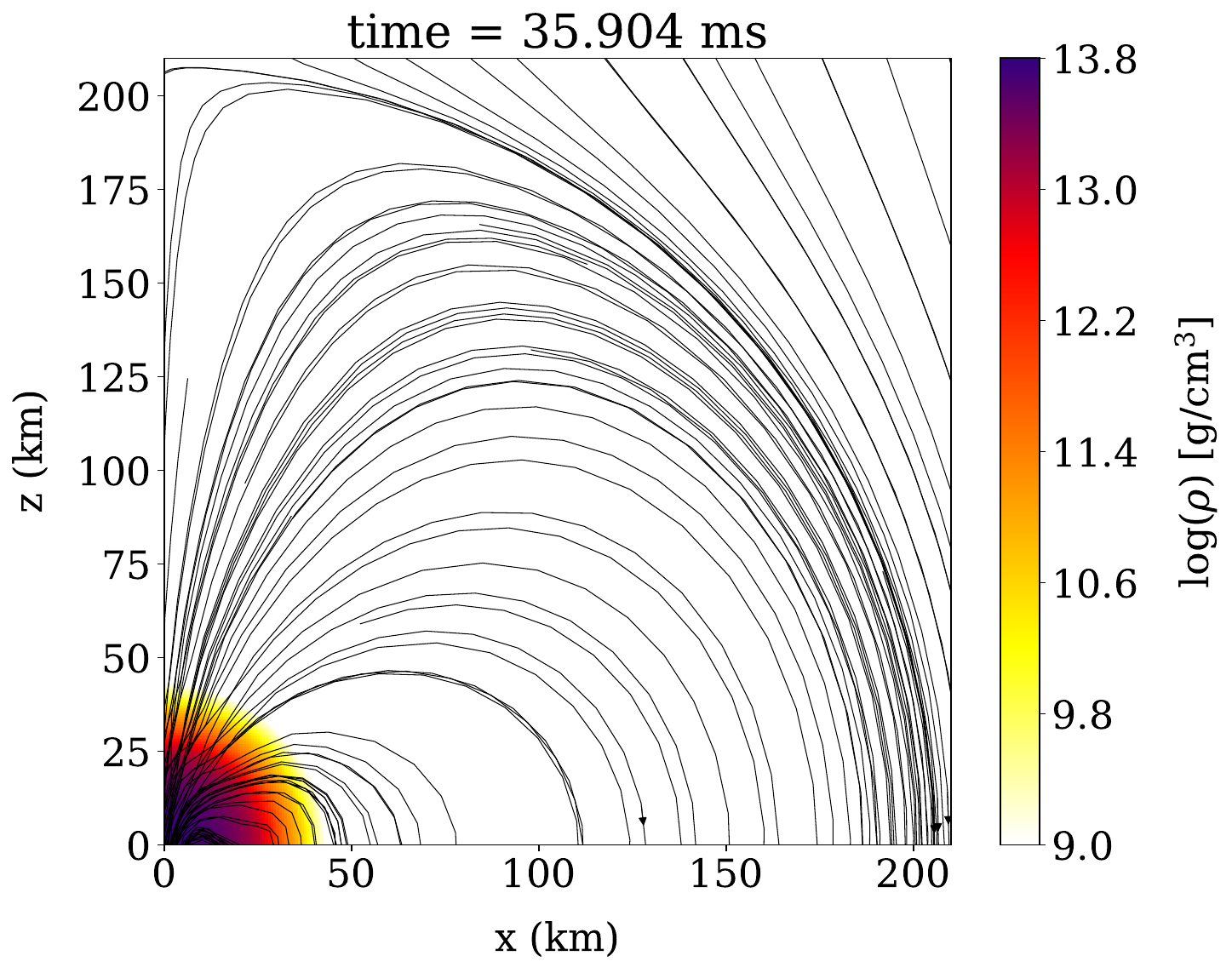}
    \end{subfigure}
    
     \caption{\small 
        We show two-dimensional slices in the \( (x, z) \) plane of the rest-mass density and magnetic field lines with a resolution of \( \Delta x = 0.5\,M_\odot \) at the initial time \( t = 0\,\mathrm{ms} \), and at \( t = 4.311\,\mathrm{ms} \), \( t = 8.252\,\mathrm{ms} \), and the final time \( t = 35.904\,\mathrm{ms} \), respectively. The upper left panel shows a non-rotating star at $t = 0 \,\text{ms}$ with an initial magnetic field extending beyond the stellar surface. The other panels show the evolution of the magnetic field and density at different times. The magnetic field configuration remains stable over time as the star evolves.
    }
    \label{fig:Extended_B_plots}
\end{figure}

Fig.~\ref{fig:Extended_B_plots} shows two-dimensional cuts in the $(x,z)$ plane of the rest-mass density and magnetic field lines at the initial time $t = 0$ ms, $t = 4.311$ ms, $t = 8.252 $ ms, and the final time $t = 35.904$ ms, respectively. The upper left panel of Fig.~\ref{fig:Extended_B_plots} shows that, at the initial time, the magnetic field lines extend both inside the star and beyond its surface. As the evolution proceeds, it is evident that the code is capable of accurately reproducing the stable evolution of the star. 
As explained earlier, the magnetic field is initially not in equilibrium with the star, and this is also evident in the top two panels of Fig.~\ref{fig:Extended_B_plots}. The star gradually approaches a quasi-equilibrium state after approximately \( 8\,\mathrm{ms} \).

\section{Conclusion}
\label{sec:Conclusion_sec}
In this work, we present a new general relativistic resistive magnetohydrodynamics module in the GPU-accelerated GRMHD code \texttt{GRaM-X}. This module extends the capabilities of \texttt{GRaM-X} beyond ideal MHD.
\texttt{GRaM-X} is optimized for simulating relativistic plasmas in a dynamical spacetime. Its GPU-based design provides us with the advantage of significant speedups and cost-effectiveness compared to conventional CPU-based implementations, enabling high-resolution and large-scale studies of complex astrophysical phenomena~\cite{Shankar2023, shankar20253d, halevi2025black, schnauck2025gravitational}. The resistive MHD framework overcomes the key limitations of the ideal MHD approximation by systematically incorporating electrical resistivity, which is necessary to capture fundamental processes like Ohmic dissipation and magnetic reconnection in high-energy, low-density astrophysical plasmas. These plasma environments play a crucial role in neutron star mergers, accretion disks around black holes, magnetospheres, and relativistic jets. As a result, the framework bridges the gap between simplified models and physically complete descriptions of relativistic plasmas.
We build the GR-RMHD module by coupling Maxwell's equations to the hydrodynamic ones, and implement an IMEX-RK2 scheme which ensures numerical stability even in regimes where source terms are stiff~\cite{Pareschi2005, Palenzuela2009}. Furthermore, we incorporate a reasonably robust one-dimensional iterative solver for the conservative to primitive transformation in the presence of stiff electric fields, which is essential to ensure the stability of the code.

We validate the stability and robustness of the new GR-RMHD module through a set of well-known tests. These tests included one-dimensional shocktubes to verify the accuracy in dealing with shocks and discontinuities; large-amplitude circular Alfv\'{e}n waves to demonstrate the ability of smooth wave propagation in the resistive regime; a two-dimensional cylindrical explosion to investigate multidimensional shock interactions; and non-rotating magnetized stars to test the coupling of magnetic fields with self-gravitating matter in a stable configuration~\cite{Palenzuela2009, Dionysopoulou2013, Bucciantini2012}.
Comparison of the results with the analytical solution (or the ideal MHD solution in the high conductivity limit) shows that in all these tests, the code produces consistent and physically meaningful results over a wide range of conductivities, with or without the presence of strong discontinuities and shocks.

Beyond these validation studies, the significance of this new module lies in the new scientific opportunities it enables. Adding resistivity to \texttt{GRaM-X} allows us to simulate processes that are inaccessible in ideal MHD models, including reconnection events in neutron star mergers, magnetic energy dissipation in the compact object's magnetosphere, and resistivity-driven dynamics in accretion flows. These processes play a fundamental role in the generation of observable signals, including the electromagnetic counterparts of gravitational wave events, making resistive GRMHD an essential tool for interpreting astrophysical multi-messenger observations~\cite{palenzuela2013linking, Dionysopoulou2013}.

Although the present development marks an important step in the extension of \texttt{GRaM-X} to the resistive MHD regime, further improvements are required to allow fully realistic astrophysical simulations. In particular, carrying out more realistic simulations of high-energy astrophysical systems requires the use of microphysical EoS, including tabulated ones. In order to use such EoS in simulations as well as to deal with more extreme simulation regions—for example, those with higher Lorentz factors (\( \gtrsim 2 \))—it is necessary to implement more robust conservative to primitive solvers than those currently implemented in \texttt{GRaM-X}. We have already begun development and integration of these methods, and this will be a primary focus in the next year such that we can apply resistive \texttt{GRaM-X} to a broader range of scenarios with higher physical fidelity.

\section*{Acknowledgments}
\label{sec:ack}

SA and PM acknowledge funding through NWO under grant No. OCENW.XL21.XL21.038. ES acknowledges the support of the Natural Sciences and Engineering Research Council of Canada (NSERC). Research at Perimeter Institute is supported in part by the Government of Canada through the Department of Innovation, Science and Economic Development and by the Province of Ontario through the Ministry of Colleges and Universities. We thank SURF (\url{www.surf.nl}) for the support in using the National Supercomputer Snellius. This research used resources of the Oak Ridge Leadership Computing Facility at the Oak Ridge National Laboratory, which is supported by the Office of Science of the U.S. Department of Energy under Contract No. DE-AC05-00OR22725. Some of the simulations were carried out on OLCF’s Frontier using the allocation AST191.

\bibliographystyle{apsrev4-2}
\bibliography{references}

\end{document}